\documentclass[usenatbib,useAMS,a4paper]{mn2e}

\usepackage{graphicx,times}

\newcommand{\Msun}{\ensuremath{M_{\odot}}}

\newcommand{\Mgtwo}{\ensuremath{{\rm Mg}_2}}

\newcommand{\aFe}{\ensuremath{\alpha/{\rm Fe}}}

\newcommand{\ZH}{\ensuremath{Z/{\rm H}}}

\newcommand{\ur}{\ensuremath{u\!-\!r}}

\title[Environment and self-regulation in galaxy formation]
{Environment and self-regulation in galaxy formation}

\author[D.~Thomas et al.] {
\parbox[h]{\textwidth}{Daniel Thomas$^{1}$, Claudia Maraston$^{1}$, Kevin Schawinski$^{2,3}$, Marc Sarzi$^4$, Joseph Silk$^5$}
\vspace*{8pt}\\ 
$^1$Institute of Cosmology and Gravitation, University of Portsmouth, Dennis Sciama Building, Burnaby Road, Portsmouth, PO1 3FX, UK\\
$^2$Department of Physics, Yale University, New Haven, CT 06511,
USA\\
$^3$Yale Center for Astronomy and Astrophysics, Yale University,
P.O. Box 208121, New Haven, CT 06520, USA\\
$^4$Center for Astrophysics Research, University of Hertfordshire, College Lane, Hatfield, Herts AL10 9AB, UK\\
$^5$Astrophysics, University of Oxford, Keble Road, Oxford, OX1 3RH, UK}

\date{Accepted 2010 January 26.  Received 2010 January 19; in original form 2009 November 27}

\pagerange{\pageref{firstpage}--\pageref{lastpage}}

\pubyear{2010}

\begin{document}

\maketitle

\label{firstpage}

\begin{abstract}
The environment is known to affect the formation and evolution of galaxies considerably best visible through the well-known morphology-density relationship. It is less clear, though, whether the environment is equally important at a given galaxy morphology. In this paper we study the effect of environment on the evolution of early-type galaxies as imprinted in the fossil record by analysing the stellar population properties of 3,360 galaxies morphologically selected by visual inspection from the Sloan Digital Sky Survey in a narrow redshift range ($0.05\leq z\leq 0.06$). The morphological selection algorithm is critical as it does not bias against recent star formation. We find that the distribution of ages is bimodal with a strong peak at old ages and a secondary peak at young ages around $\sim 2.5\;$Gyr containing about 10 per cent of the objects. This is analogue to 'red sequence' and 'blue cloud' identified in galaxy populations usually containing both early and late type galaxies. The fraction of the young, rejuvenated galaxies increases with both decreasing galaxy mass and decreasing environmental density up to about 45 per cent, which implies that the impact of environment increases with decreasing galaxy mass. The rejuvenated galaxies have lower \aFe\ ratios than the average and most of them show signs of ongoing star formation through their emission line spectra. All objects that host AGN in their centres without star formation are part of the red sequence population. We confirm and statistically strengthen earlier results that luminosity weighted ages, metallicities, and \aFe\ element ratios of the red sequence population correlate well with velocity dispersion and galaxy mass. Most interestingly, however, these scaling relations are not sensitive to environmental densities and are only driven by galaxy mass. We infer that early-type galaxy formation has undergone a phase transition a few billion years ago around $z\sim 0.2$. A self-regulated formation phase without environmental dependence has recently been superseded by a rejuvenation phase, in which the environment plays a decisive role possibly through galaxy mergers and interactions.
\end{abstract}



\section{Introduction}
\label{sec:intro}
The environment is known to be a major driver in the formation and evolution of galaxies. Its influence is best visible through the well-known morphology-density relationship, according to which
early-type galaxies and morphologically undisturbed galaxies are
preferentially found in high density environments and vice versa
\citep{Oemler74,MS77,Dressler80,PG84}. The existence of this
environmental footprint is well established, and has been repeatedly
confirmed both in the local
\citep{Tranetal01,Gotoetal03,HP03,KR05,Parketal07,SADR06,Holdenetal07,vanderWeletal07,Bamfordetal09} and
intermediate redshift universe
\citep{Dressleretal97,Fasanoetal00,Treuetal03,Smithetal05,vanderWeletal07,Panetal09b,Tasca09}. While the
exact mechanisms responsible for this effect are still debated
\citep[e.g.][]{KS01,Baloghetal02,Treuetal03}, its mere existence is in
good agreement with the idea of hierarchical structure formation based
on cold dark matter cosmology \citep{Blumenthaletal84,Davisetal85}.

In contrast, it is less clear whether the environment is equally
important at a given galaxy morphology. There is still major
controversy about whether the formation and evolution of the most
massive and morphologically most regular galaxies in the universe,
i.e.\ early-type galaxies, are affected by environmental
densities. In \citet[][hereafter T05]{Thomasetal05} studying the local early-type galaxy population (what we call the 'astro-archaeology approach') we find significant evolution as a function
of environment in agreement with a large number of investigations in the literature \citep{Traetal00b,Lonetal00,Poggetal01b,Kunetal02b,TF02,Procetal04,Denicoloetal05,Mendesetal05,Colloetal06,Schawetal07a,Sanchetal06,RSO07,Annetal07,delaRosaetal07}. In
summary, early-type galaxies in lower densities appear younger by
$1-2\;$Gyr at given galaxy mass.

These differences should become even more pronounced when probing earlier epochs. As discussed in the review by \citet{Renzini06}, however, studies of the intermediate redshift early-type galaxy population do not confirm this picture. The same colour evolution for both field and cluster galaxies is found in recent redshift surveys
\citep{Wolfetal03,Belletal04,Kooetal05b,Tanakaetal05,Bundyetal06}. This negligible influence from the environment is further supported by investigations of the fundamental plane of ellipticals, star formation rates, luminosity function etc. back to redshift $z\sim 1$, that again find similar evolution in clusters and in the field
\citep{Treuetal01,vDE03,vdWeletal05,dSAetal05,Jorgetal06,vDvdM06,dSAetal06,FHB06,Strazzetal06,Panetal09b}. Finally, \citet{Wakeetal05} study clusters at $z\sim 0.3$ with a large range of X-ray properties and also find a relatively small influence of the environment on the colour-magnitude relation.

With this paper we utilise the SDSS database to check the result found in T05 via an increase of the sample size by more than an order of
magnitude. The Sloan Digital Sky Survey \citep[SDSS,][]{Yorketal00} provides the opportunity to explore huge homogeneous samples of early-type galaxies in the nearby universe, comprising several ten thousands of objects, so that a statistically meaningful investigation of the stellar population parameters of galaxies and their dependence on environment can be attempted. We analyse the stellar population parameters luminosity-weighted age, metallicity, and \aFe\ element ratio of 3,360 early-type galaxies drawn from the SDSS in a narrow redshift range ($0.05\leq z\leq 0.06$). A major strength of our approach is the application of strict morphological selection criteria based on visual inspection of 48,023 galaxies. We confirm the basic result of T05 as far as 'archaeological downsizing' and the major formation epochs of early-type galaxies are concerned. However, different from T05, the latter turn out to be insensitive to changes of the environment for the bulk of the early-type galaxy population.

The paper is organised as follows. In Section~2 we describe in brief
the construction of the galaxy catalogue, and the method to derive
stellar population parameters. The results on stellar population scaling relations are presented in Section~3, galaxy formation epochs are discussed in Section~4. The paper concludes with discussion of the results and literature comparison in Sections~5 and~6.

\section{Data sample and analysis}
\label{sec:data}
The sample utilised here is part of a project called MOSES:
\textbf{MO}rphologically \textbf{S}elected \textbf{E}arly-types in
\textbf{S}DSS. We have collected a magnitude limited sample of 48,023 galaxies in the redshift range $0.05\leq z\leq 0.1$ with apparent $r$-band magnitude brighter than 16.8 from the SDSS Data Release 4
\citep{Adeletal06}. Model magnitudes are used throughout this work. The most radical difference with respect to other
galaxy samples constructed from SDSS is our choice of {\rm purely
morphological} selection of galaxy type. All galaxies in this sample have been visually inspected and classified by eye into early- and late-type morphology. We have already used this approach in \citet{Schawetal07b}, and more details on the morphological selection and other sample characteristics can be found there. In the following, we provide a brief summary with focus on the aspects that are most relevant for the present work.

\subsection{Morphological selection}
We {\em visually} inspected the 48,023 objects and divided the sample
in 31,521 late-type (spiral arms, clear disc-like structures) and
16,502 early-type (roundish, elliptically shaped)
galaxies. Hence, 34 per cent of the objects are classified as 'early-type'. This is in good agreement with the typical value found in average environments (see references in the Introduction). For the present study we aim to minimise contamination by late-type galaxies, hence our relatively strict selection criteria yield a conservative early-type galaxy fraction that is slightly smaller than the recent determination based on SDSS by \citet[$\sim 40$ per cent,][]{vanderWeletal07} but in good agreement with other determinations in the literature \citep[see][and references therein]{Bamfordetal09}.

Elliptically shaped objects with irregular structures,
potentially indicative of tidal tails originating from recent
galaxy-galaxy interaction/merger events, were included in the 'early-type'
category. The principal aim was to separate spiral and disc-like
galaxies from early-types. The major advantage of this strategy is that our sample is not biased against star forming elliptical galaxies. No further selection criteria based on stellar population properties such as colour, spectral shape, or concentration index have been applied. Indeed, this catalogue differs most from previous SDSS-based early-type galaxy catalogues \citep[e.g.,][]{Beretal03a} through its inclusion of (typically low-mass) objects with unusually blue colours. We have extended this approach recently through the Galaxy Zoo project \citep{Linetal08}, which uses several ten thousands of volunteers on the internet to classify galaxies through a web interface. Among other projects, these classifications have led to the construction of a large sample of blue early-type galaxies (Schawinski et al 2009b; see also Kannappan et al 2009)\nocite{Schawetal09b,2009AJ....138..579K}.

\subsection{Environmental density}
For the estimate of the environmental density we adopt the method
developed and described in full detail in \citet{Schawetal07a} and \citet{Yoonetal08}. In a
nutshell, we calculate the local number density of objects brighter
than a certain absolute magnitude in a $6\;$Mpc sphere around the object of
interest. A Gaussian weight function centred on the object ensures to
avoid contamination from neighbouring structures, and an adjustment of the side length of the 'sphere' in redshift direction is invoked in order to account for the 'finger of God effect'. We have tested various sizes for the Gaussian weight functions and found that the final results of this work are not sensitive to the particular size of the sphere around $6\;$Mpc. Note that the results using spheres of only a few Mpc size were noisier because of the sparse sampling by SDSS fibers. The $6\;$Mpc chosen here can be considered conservative for the purposes of this work in that this choice assures the definition of low environmental density to indicate truly isolated regions.

We obtain the 3-dimensional volume density at the location
of each object, hence an estimate of the {\em local} environmental
density. This process does not constrain the overall structure in
which the object resides, but is a sophisticated number density providing a measure
of the number and proximity of galaxies around a point
in space. In other words, we cannot distinguish the
outskirts of a galaxy cluster from a group environment of the same
local density.  We decided not to include further constraints like
cross-correlation with cluster catalogues like in, e.g.,
\citet{Beretal06}, because also this method suffers from
incompleteness problems.  More generally speaking, we did not attempt to correlate our density estimates with physical structures, as we are not aiming for classifications such as 'field', group', or 'cluster'.
In any case, as discussed in Section~5, our results are overall consistent
with other SDSS studies in the literature as far as the general
influence of the environment is concerned.

\subsection{Kinematics and Lick line index measurements}
The kinematics of gas and stars are determined for all 48,023 objects,
using the code GANDALF developed by \citet{Sarzietal06}. Stellar population and
emission line templates are fitted {\em simultaneously} to the galaxy
spectrum. The outcome are emission line fluxes and kinematics of the
gas. On the stellar population side, we obtain line-of-sight velocity
dispersions based on the pixel fitting method of
\citet{CE04} with a typical error of $0.023\;$dex in $\log\sigma$. The error distribution deviates from a Gaussian with a sharp cut-off at $0.01\;$dex and a tail extending to larger errors of $\sim 0.1\;$dex. Through the subtraction of the emission line spectrum from
the observed one, we get the clean absorption line spectrum free from
emission line contamination. This spectrum is used for any further
analysis of absorption features. The stellar template spectrum is not
used for scientific analyses.

On each spectrum, we measure the 25 standard Lick absorption line indices \citep{Woretal94,WO97} following the most recent index
definitions of \citet{Traetal98}. For this purpose the spectral
resolution is reduced to the wavelength-dependent Lick resolution
\citep{WO97}. The measurements are then corrected for velocity
dispersion broadening. The correction factor is evaluated using the
best fitting stellar template and velocity dispersion obtained
previously. Errors are determined by Monte Carlo simulations on each
spectrum individually based on the signal-to-noise ratios provided by
the SDSS.

It should be noted that an important step in the measurement of Lick
indices cannot be undertaken. Ideally, Lick standard stars need to be
observed with the same instrument set-up, in order to determine (and
correct for) offsets caused by the shape of the spectral energy
distribution. This is in principle necessary, as the Lick/IDS system
is not based on flux-calibrated spectra. Such calibration stars at sufficiently high signal-to-noise are not available in SDSS. However, it ought to be expected that the shape of the spectrum is of little
importance for absorption indices where line feature and pseudo-continuum windows are narrow and close enough in wavelength. This condition is actually fulfilled for most Lick indices
with the pseudo-continuum bands being placed directly adjacent to the
line feature itself. A detailed analysis of Lick standard stars extracted from the SDSS archive suggests that deviations from the true Lick system are indeed negligible \citep{Carson07}.

\subsection{Stellar population parameters}
We use the stellar population models of \citet[][hereafter
TMB]{TMB03a,TMK04} to derive luminosity-weighted ages, metallicities, and \aFe\ ratios from the Lick absorption line indices by means of the minimised $\chi^2$ technique introduced by \citet{PS02}.
The main motivation for this choice is that the $\chi^2$ fitting is very efficient and fast, and can easily be applied to large data sets like SDSS. A further advantage is that it considers a large wavelength range of the spectrum and hence gives a better estimate of the real average population. Finally, the consideration of a larger set of absorption features increases the signal-to-noise ratio.
Errors are obtained from the probability function based on the $\chi^2$ distribution. In practice, we search for the combination of those three parameters that minimise the $\chi^2$ between model and observational data for all 25 indices observed. This is different from the approach followed in T05, where only a selection of well calibrated indices is considered to constrain the stellar population parameters directly rather than in a statistical fashion. It has been shown, however, that these two orthogonal methods yield consistent results \citep{TD06}.

We follow an iterative approach. In the first step we fit all 25
indices. We compute the probability distribution for different values
of $\chi^2$ at its minimum using the incomplete $\Gamma$
function. This yields the probability $P$ that the observed $\chi^2$
for a correct model should be less than the value $\chi^2$ obtained in
the fit. If $P\geq 0.999$ we deem the fit unacceptable. In this case
we discard the index with the largest $\chi^2$ from the fitting
procedure, re-do the fit, and re-calculate $P$. This procedure is
iterated until $P<0.999$. The indices Ca4455 and NaD were discarded in more than 10 per cent of the cases, and were hence generally
excluded. Their exclusion is further supported by the fact that these indices are not well calibrated \citep{Maretal03,TMB03a}. In more than half of the cases (65 per cent), no further
index was removed (hence all 23 indices were used). In 20, 8, and 4
per cent of the objects 1, 2, and 3 indices were removed,
respectively. In other words, for 97 per cent of all objects, at least
20 indices were included in the fit.

The average errors in log(age), metallicity, and [\aFe] ratio are $0.10\;$dex, $0.07\;$dex, and
$0.06\;$dex, respectively. The error distributions are well approximated by Gaussian functions with standard deviations of 0.04, 0.02, and $0.03\;$dex, respectively. Finally it should be noted that size of the SDSS fibers used for the
spectroscopy is relatively large (3 arcsec). The stellar population properties
discussed here therefore refer to the global stellar populations in the galaxies
roughly within or only slightly below the half-light
radius. As early-type galaxies are known to have metallicity gradients \citep[e.g.][]{DSP93} the effect of these when using light-average quantities is that we slightly underestimate the true age of the stellar population, simply because the metal poor halo provides blue light which is interpreted as a slightly younger population. We expect this effect to be small, though, because the amount of metal-poor stars in luminous red galaxies in SDSS is of the order of only a few per cent by mass \citep{Marastonetal09}.
Moreover, the \aFe\ ratio is not affected by this as early-type galaxies exhibit no significant gradient in \aFe\ within the effective radius \citep{Mehetal03,Annetal07,Rawleetal08}. Note also that no gradient in age is typically found in local early-type galaxies \citep{Sagetal00,Tamuraetal00,Mehetal03,Wuetal05,Annetal07,Rawleetal08}.

\subsection{Final selection}
\begin{figure*}
\includegraphics[width=0.49\textwidth]{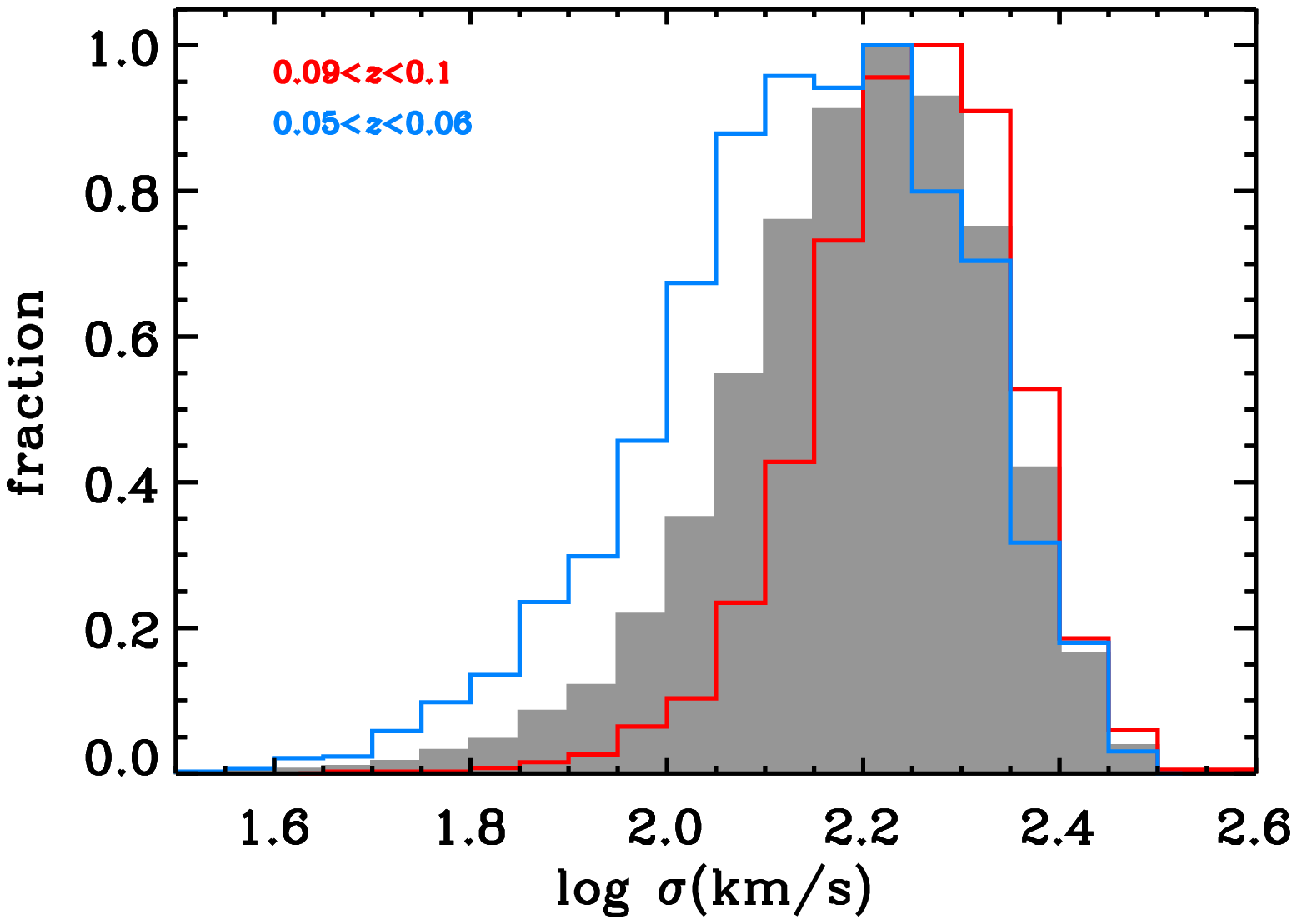}
\includegraphics[width=0.49\textwidth]{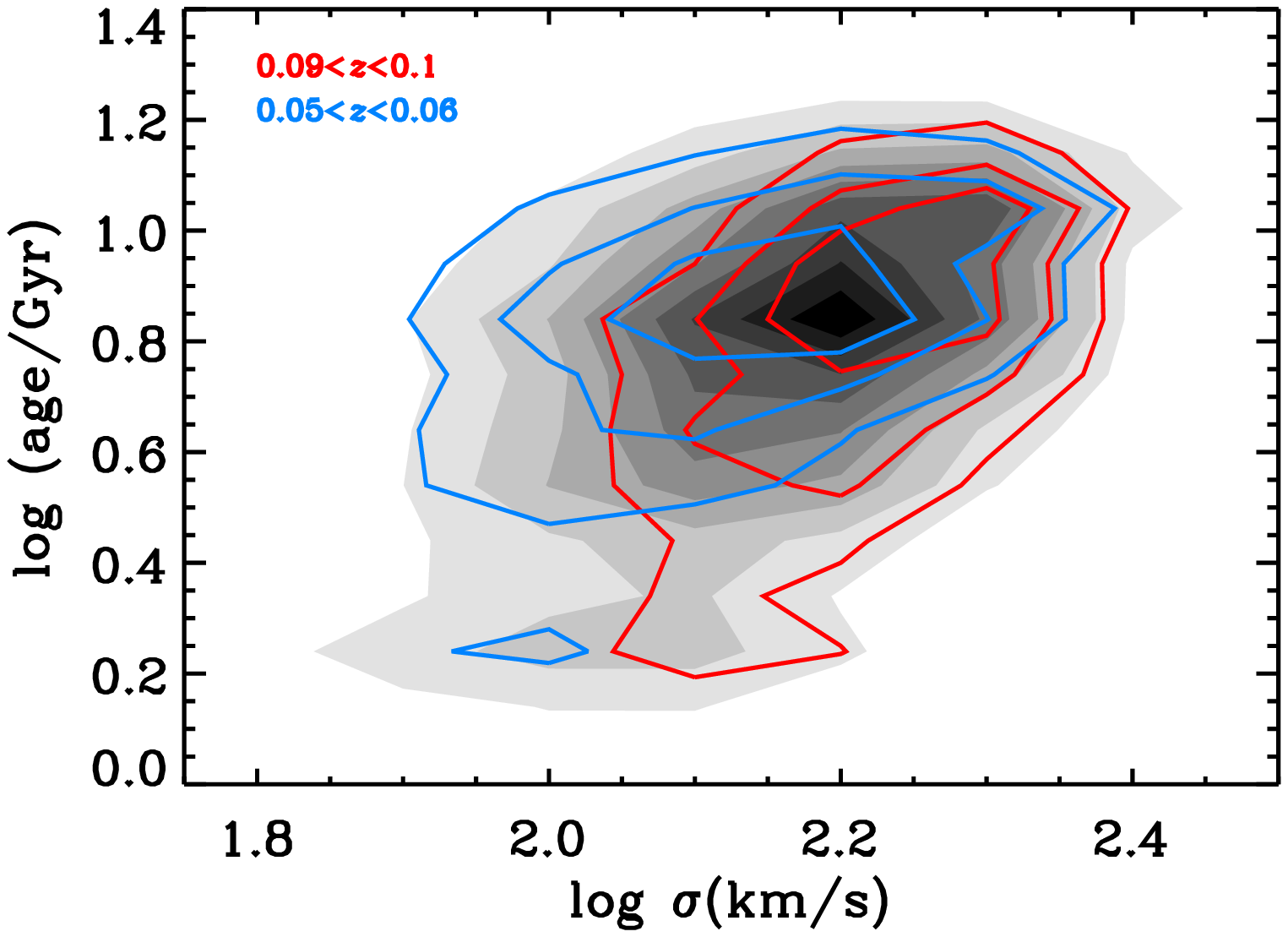}
\caption{{\em Left-hand panel:} Distribution of velocity dispersion for different sample selections in redshift. The grey shaded histogram is the entire sample of 16,502 early-type galaxies in the redshift range $0.05\leq z\leq 0.1$. Blue and red histograms show low and high redshift selections, respectively. {\em Right-hand panel:} Contours of light-averaged age as function of velocity dispersion. Colours as in left-hand panel. The final sample in this study is selected from the lowest redshift range ($0.05\leq z\leq 0.06$, blue lines).}
\label{fig:sigmaall}
\end{figure*}
We applied further selection criteria to the 16,502 early-type
galaxies in our catalogue. The redshift range sampled ($0.05\leq z\leq 0.1$) is small and corresponds to a time window of only $\sim 600\;$Myr. Still, this age difference produces significant selection effects. The left-hand panel of Fig.~\ref{fig:sigmaall} shows the distribution of velocity dispersions for low redshift ($0.05\leq z\leq 0.06$, blue line) and high redshift ($0.09\leq z\leq 0.1$, red line) selections in comparison with the entire sample (grey shaded histogram). It can be seen that despite the small range in redshift, completeness problems at the low-$\sigma$ end get worse with the inclusion of higher redshift objects. This leads to a very bad sampling of the age-$\sigma$ relationship, as shown by the right-hand panel of Fig.~\ref{fig:sigmaall}. On top of this selection effect in $\sigma$, small evolutionary effects complicate the derivation of stellar population parameters in spite of the small time window sampled \citep{Beretal06}. We therefore took a conservative approach and decided to consider only the lowest redshift bin ($0.05\leq z\leq 0.06$, blue lines in Fig.~\ref{fig:sigmaall}) providing an acceptable coverage in velocity dispersion down to $\log\sigma/{\rm km/s}\sim 1.9$. This results in a total of 3,360 early-type galaxies. This redshift range corresponds to a time interval of only $100\;$Myr at an average lookback time of $700\;$Myr adopting $\Omega_m=0.24$, $\Omega_\Lambda=0.76$, and $H_0=73\;$km/s/Mpc \citep{Tegmarketal06,Percetal07} .

\section{Stellar population scaling relations}
The characteristic mass of a galaxy population increases
with environmental density
\citep[e.g.,][]{Baldryetal06}, indicating that lower density
environments host less massive galaxies. This mass bias has to be
eliminated for a meaningful study of the influence of environment. We
therefore investigate the environmental dependence of the
correlations of the stellar population properties with velocity
dispersion $\sigma$ and dynamical galaxy mass $M_{\rm dyn}$.
From these 'stellar population scaling relations' we derive formation
histories following the method of T05.

\subsection{Previous work}
While work on the classical scaling relations of early-type galaxies
such as colour-magnitude relation, Fundamental Plane, and Mg-$\sigma$ relation has suggested relatively old ages and high formation
redshifts already almost two decades ago \citep[see review by][and references
therein]{Renzini06}, the efforts of the last decade to find
correlations between galaxy mass and the three stellar population
parameters age, metallicity, and \aFe\ ratio have succeeded only recently. Mostly because of the
well-known age-metallicity degeneracy \citep[e.g.][]{Worthey94}, the
direct determination of luminosity-weighted ages turned out to be
extremely difficult. When ages of early-type galaxies were explicitly
derived, no evidence for a correlation with mass, luminosity or
velocity dispersion was found
\citep{Jorgensen99,Traetal00b,Kuntschner00,Kunetal01,Poggetal01a,Poggetal01b,Vazetal01,TF02,Mehetal03}. As
the most noticeable exception, the study of \citet*{CRC03} hints for
the presence of an age-sigma relationship, but driven mainly by the
lowest-mass objects and with extremely large scatter.

Somewhat more convincing correlations with velocity dispersion were
found for metallicity and the \aFe\ element abundance ratio in the
studies mentioned above. In particular the latter could be further
improved \citep[T05;][]{Traetal00b,PS02,Mehetal03} after element
abundance ratio sensitive stellar population models were used
\citep[TMB;][]{Traetal00a}. Besides allowing for a quantitative
determination of element ratios, the inclusion of element ratio effects
allows for a clean definition of total metallicity and in this way
helps improving the accuracy of the age determination.

\begin{figure*}
\includegraphics[width=0.49\textwidth]{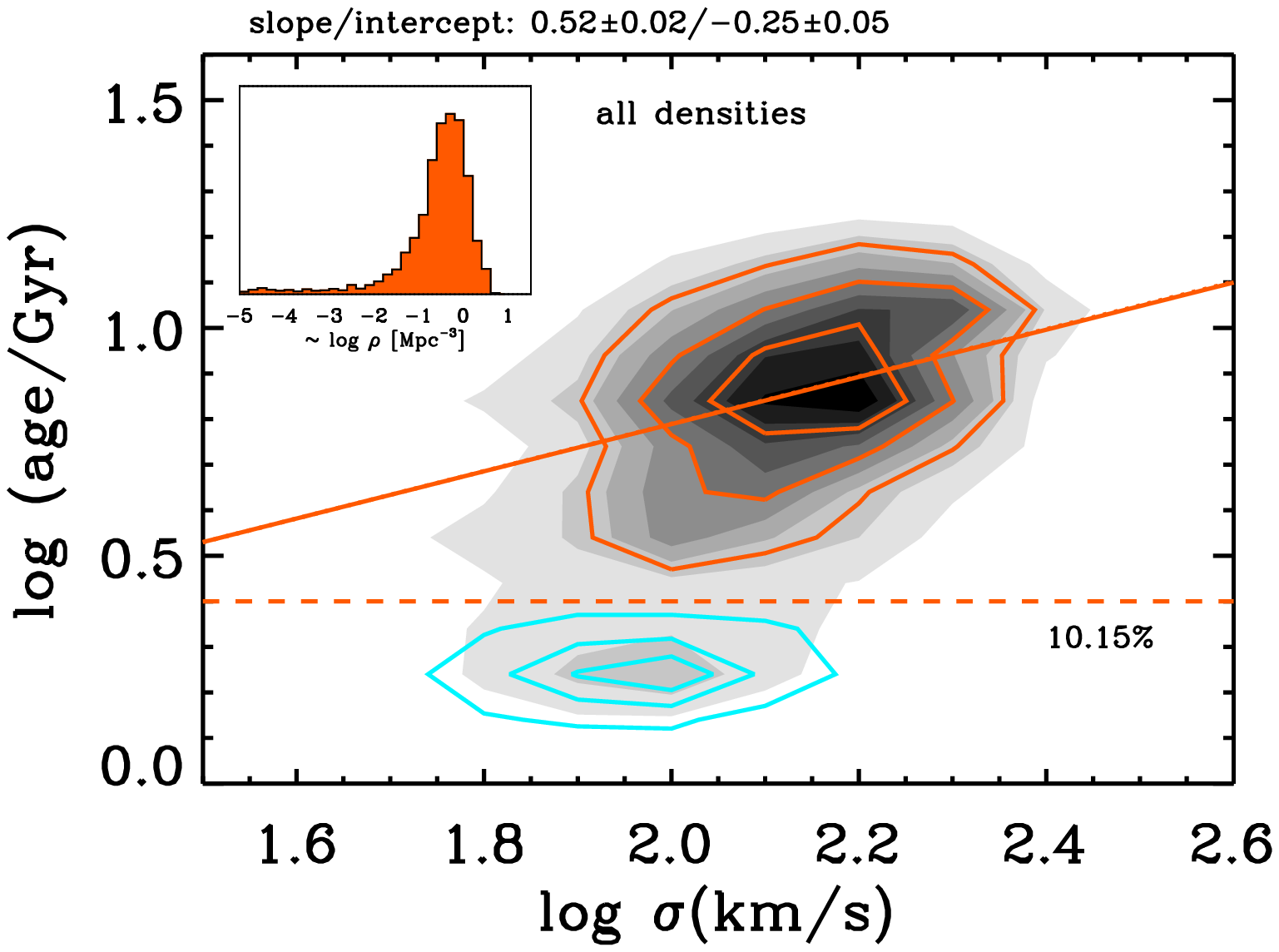}
\includegraphics[width=0.49\textwidth]{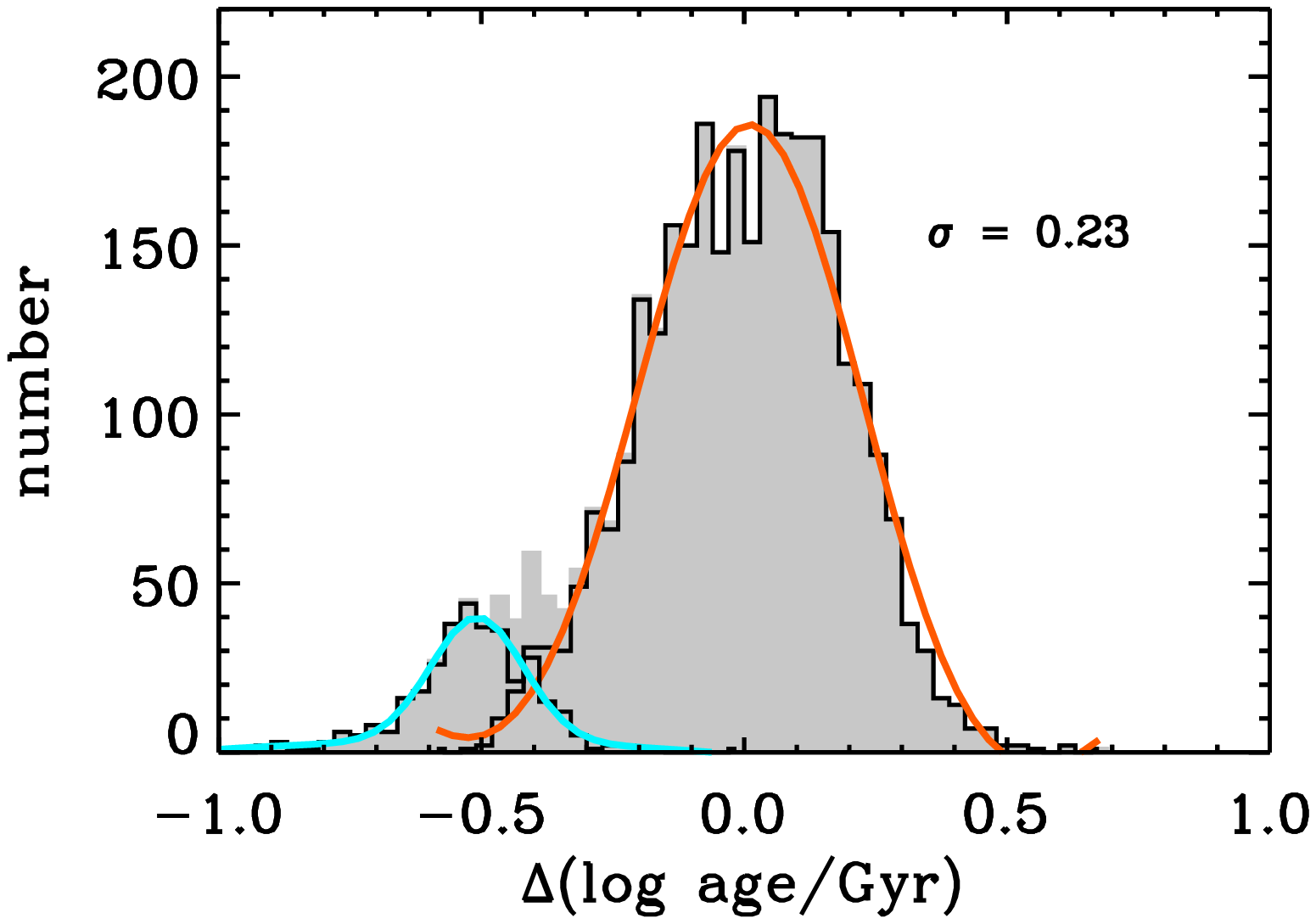}
\includegraphics[width=0.33\textwidth]{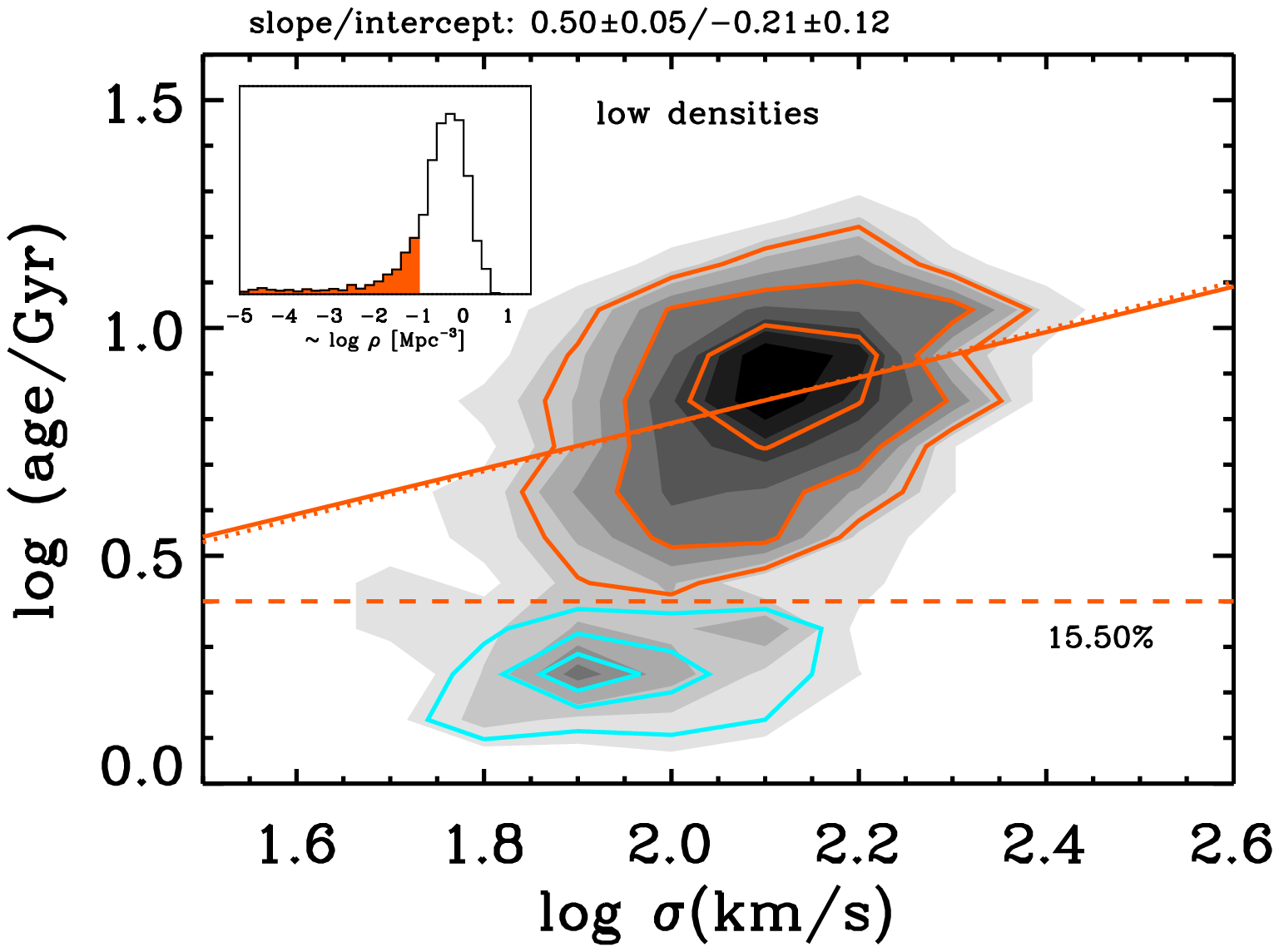}
\includegraphics[width=0.33\textwidth]{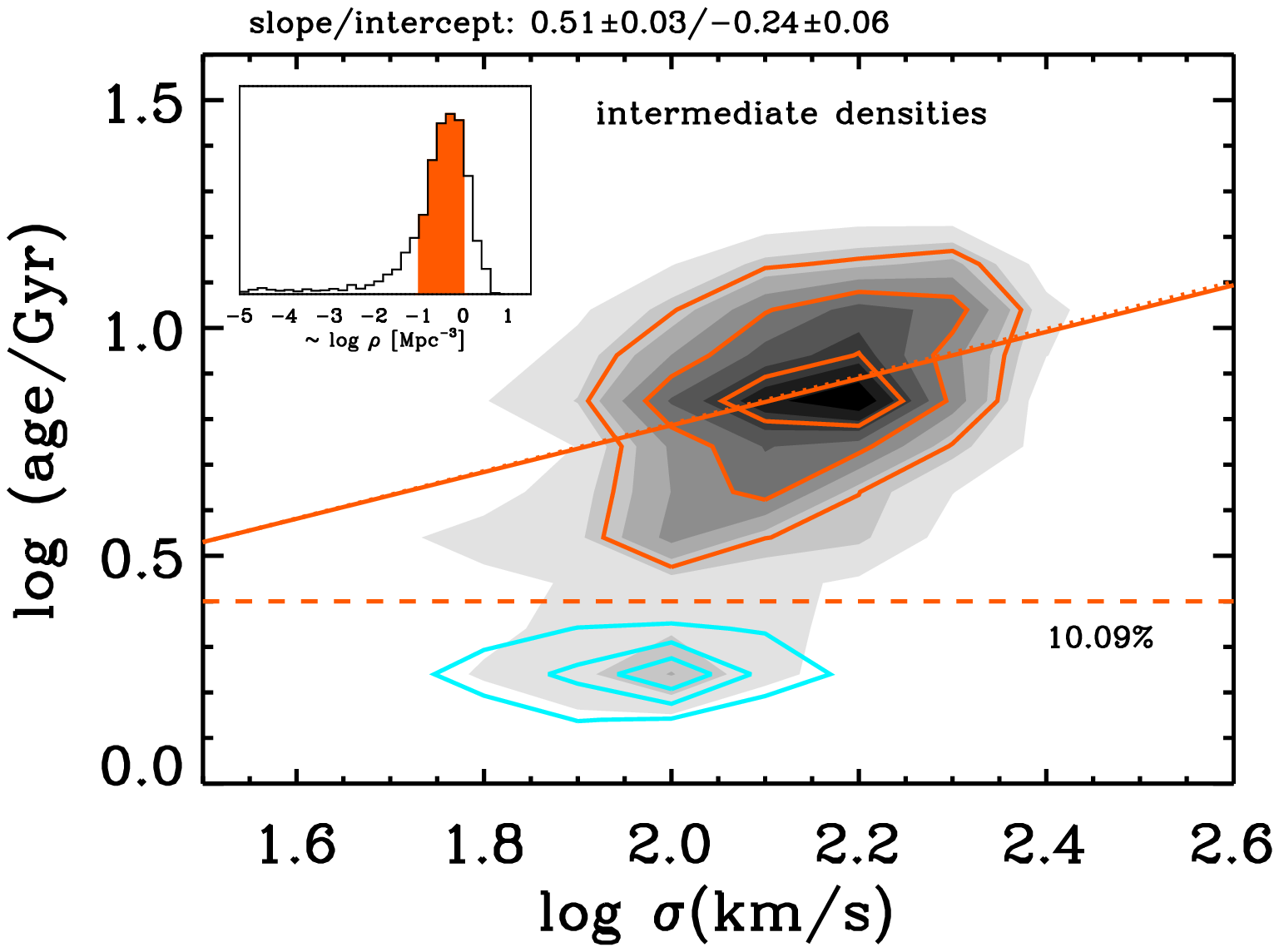}
\includegraphics[width=0.33\textwidth]{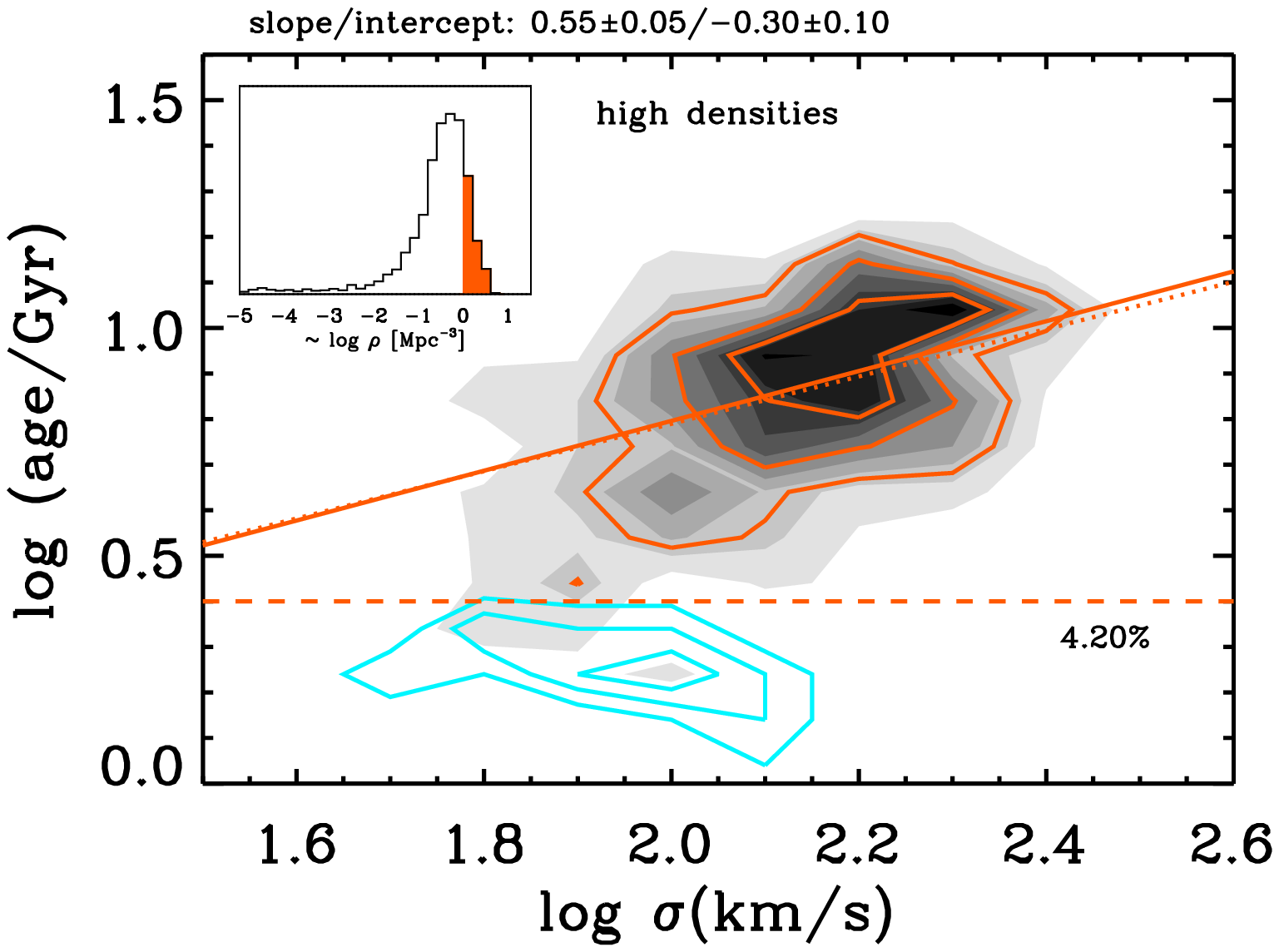}
\caption{Contour plots of the relationship between stellar velocity dispersion and luminosity-weighted age for various environmental densities as indicated by the inset histograms. The environmental density is proportional to the number density per volume, but no precise physical units are associated to it. The lowest and highest density bins contain 658 and 571 out of 3,360 galaxies, respectively. The dashed line separates an old red sequence population (orange contours) from rejuvenated objects in the blue cloud with light-averaged ages smaller than $2.5\;$Gyr (cyan contours). The fraction of this latter population is given by the label. The underlying grey contours include both populations. The solid line is a linear fit to the red sequence population, the parameters of the fit are given at the top of each panel. The dotted line is the fit for all environmental densities (parameters from top left-hand panel). Its distribution is shown by the top right-hand panel (same colour coding). The label gives the standard deviation $\sigma$ for the fit. The age-$\sigma$ relationship for the red sequence population is independent of environment, while the rejuvenation fraction increases with decreasing density.}
\label{fig:agestat}
\end{figure*}

A major obstacle, however, is that most indices depend on all three
stellar population parameters. This leads to tight correlations
between the errors of these parameters, which hampers
their determination \citep[T05;][]{Traetal00b,Kunetal01}. In particular small data sets are severely
affected by this problem. T05 follow therefore a Monte Carlo approach
and seek correlations between stellar population parameters and galaxy mass through comparison of mock galaxy samples with observational data. Through this method, well-defined relationships between all three stellar population parameters (including luminosity-weighted age) and galaxy velocity dispersion could be identified. Other recent work on extremely large data samples (mostly SDSS) basically confirm these results \citep{Nelanetal05,Beretal06,Galletal06,Clemensetal06,Jimenezetal07,Gravesetal07,2009ApJ...702.1275A}.

\begin{figure*}
\includegraphics[width=0.49\textwidth]{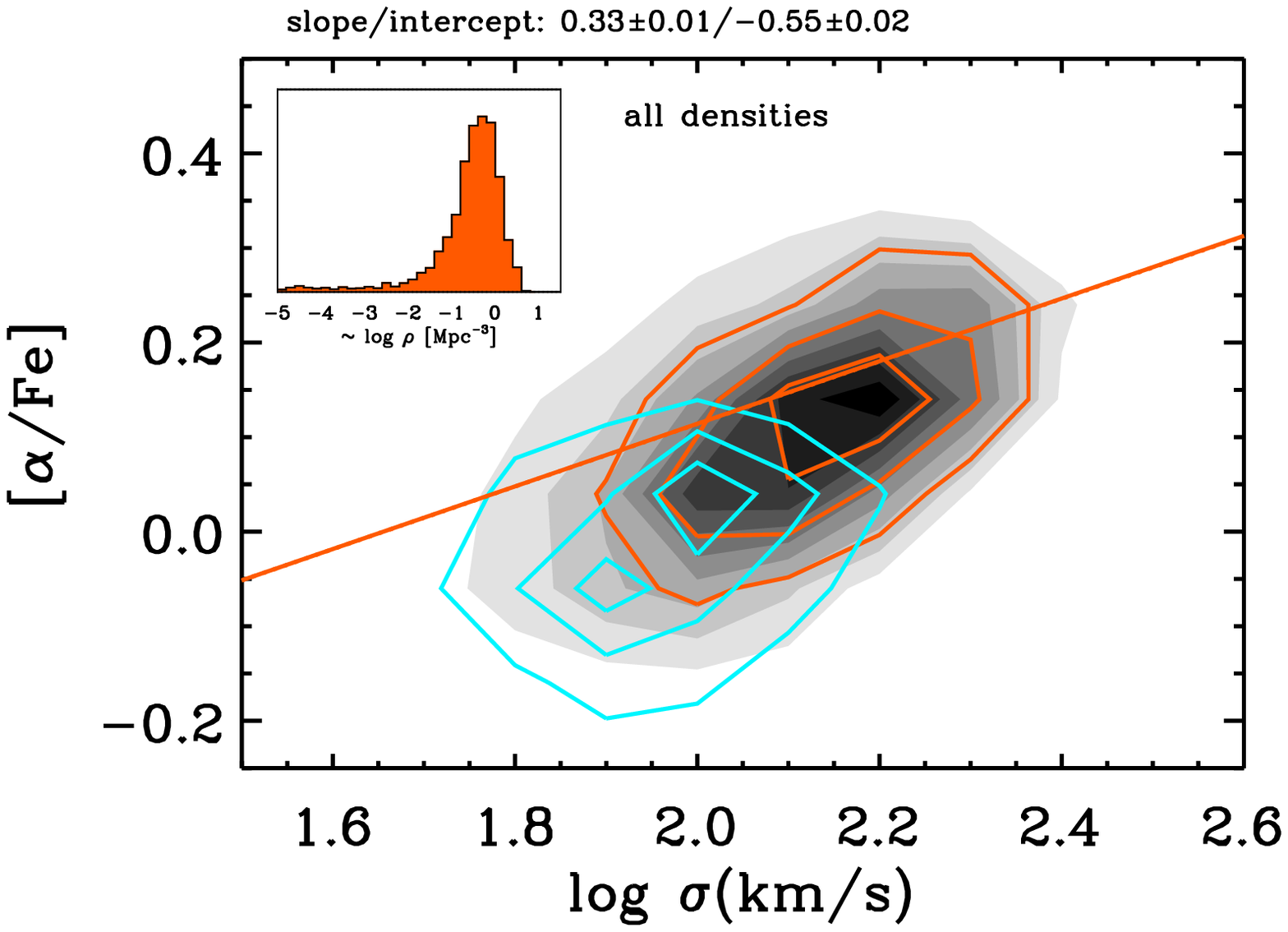}
\includegraphics[width=0.49\textwidth]{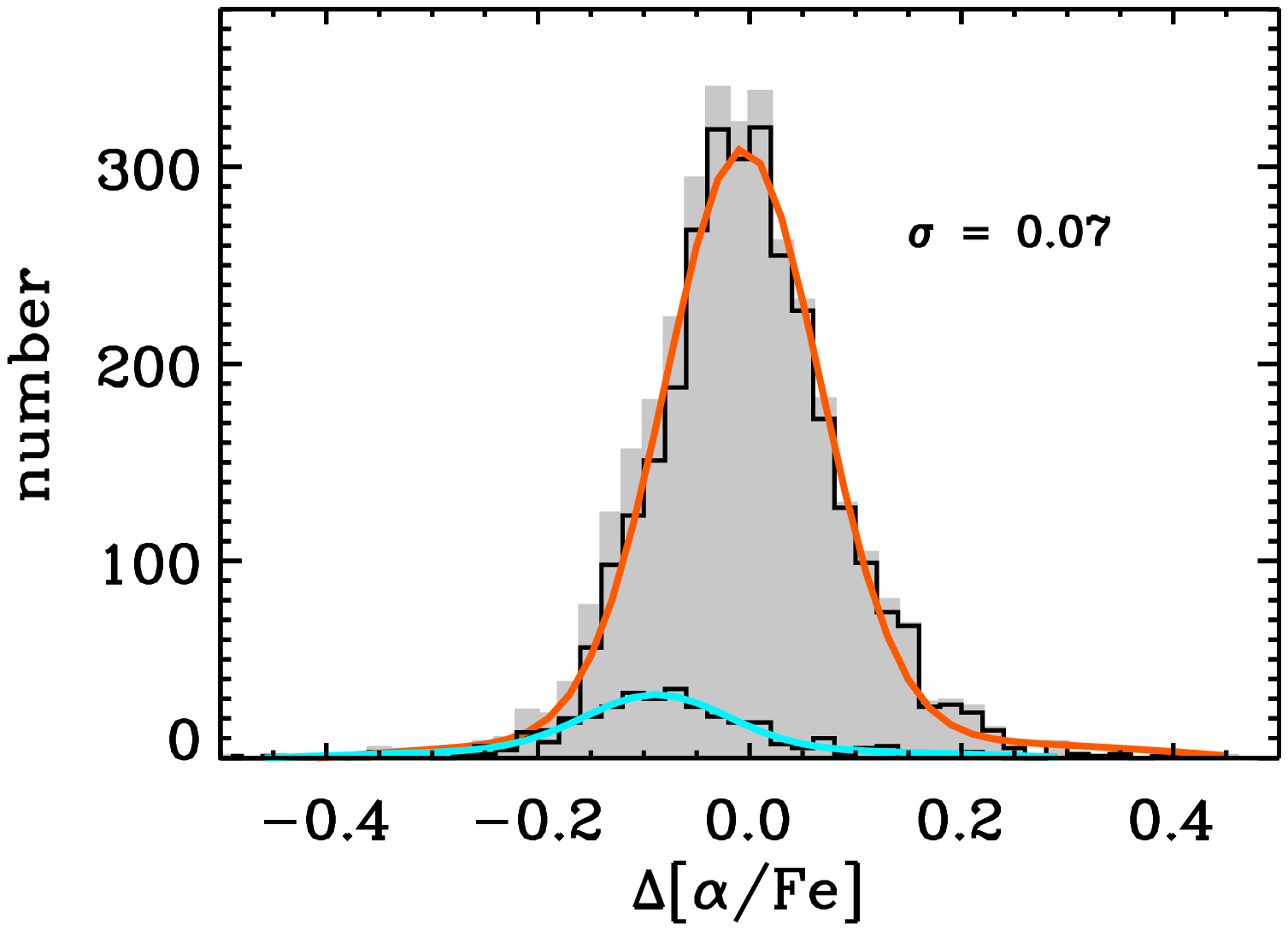}
\includegraphics[width=0.33\textwidth]{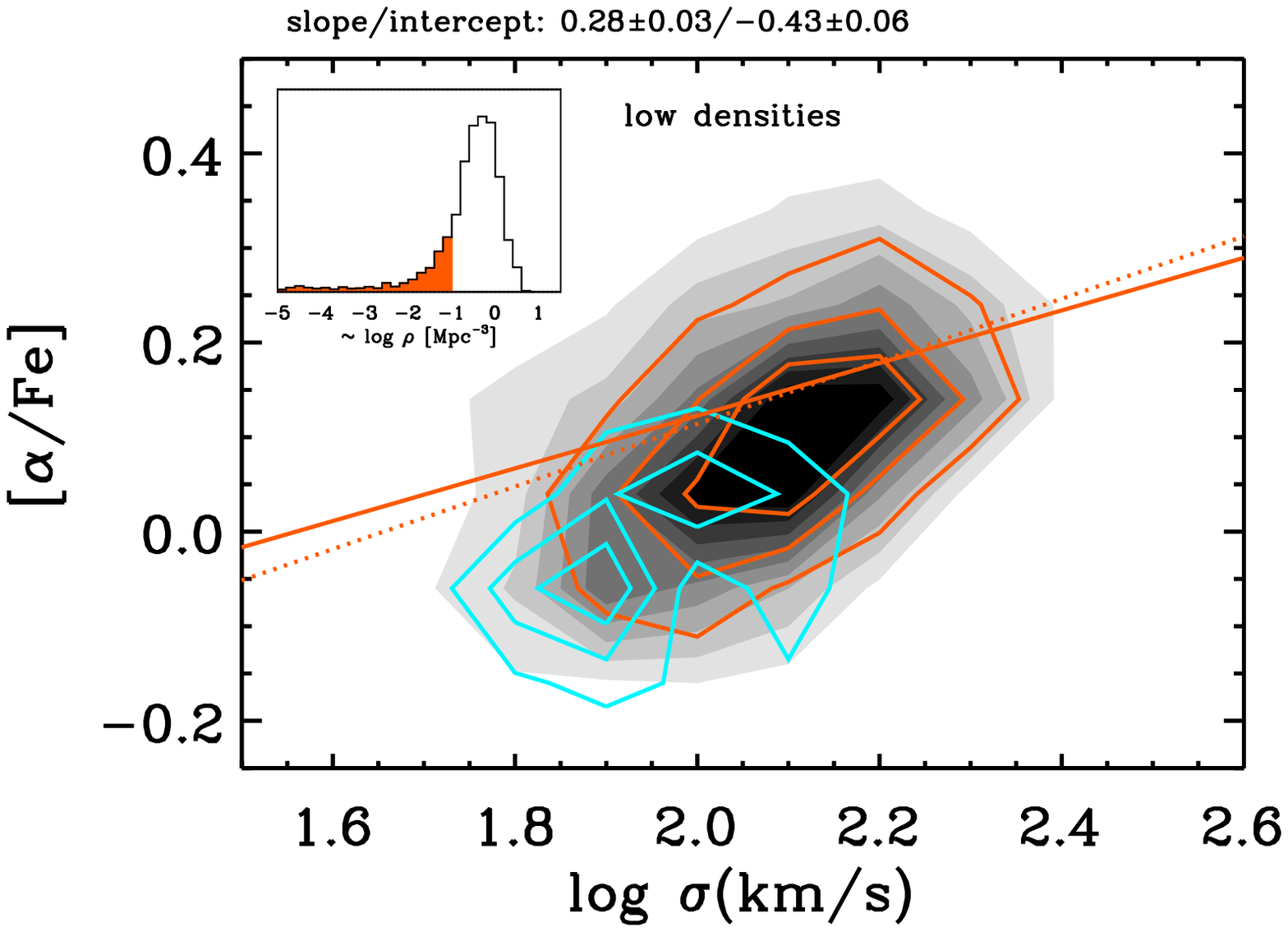}
\includegraphics[width=0.33\textwidth]{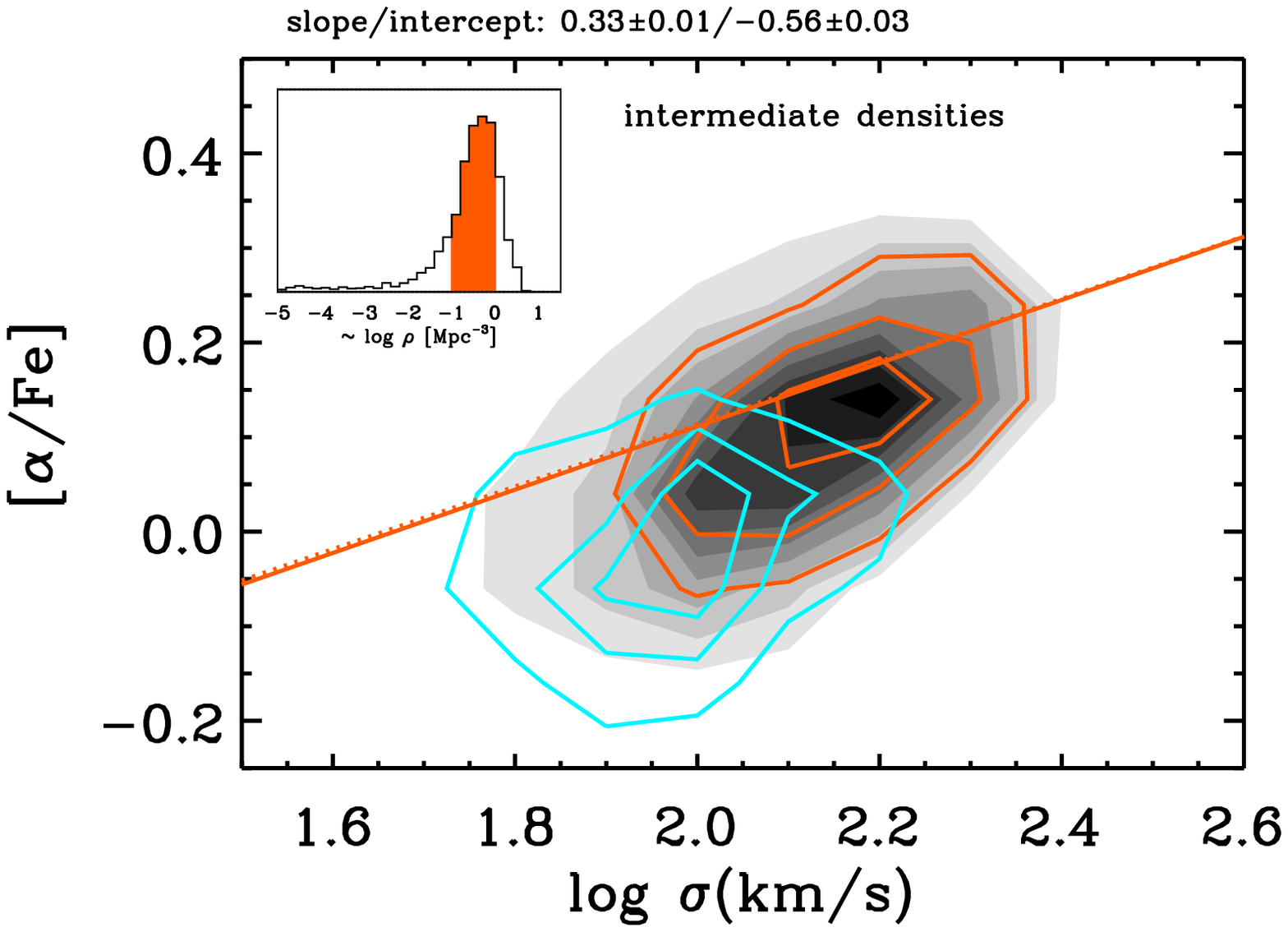}
\includegraphics[width=0.33\textwidth]{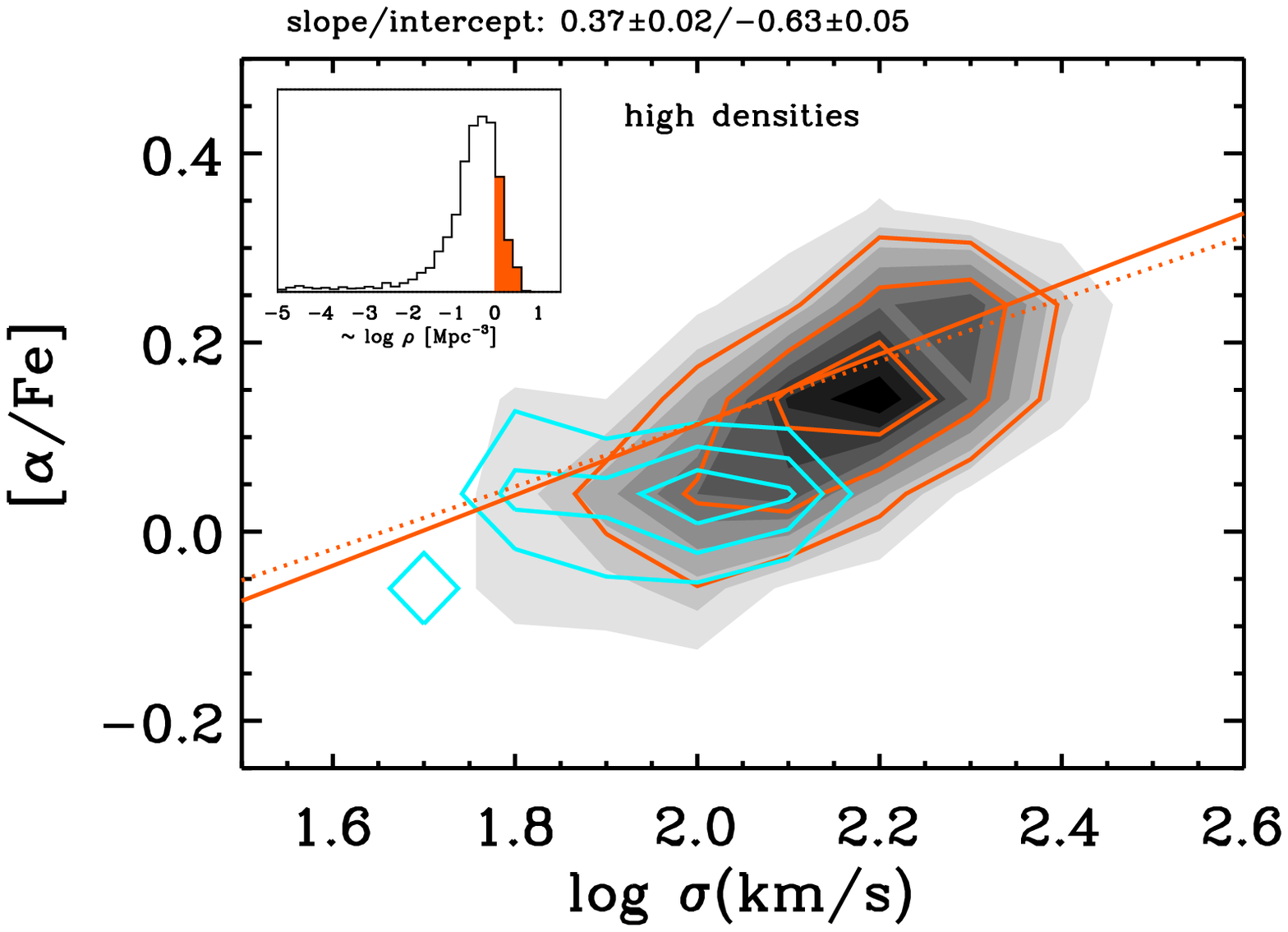}
\caption{Contour plots of the relationship between stellar velocity dispersion and \aFe\ ratio for various environmental densities as indicated by the inset histograms. The environmental density is proportional to the number density per volume, but no precise physical units are associated to it. The lowest and highest density bins contain 658 and 571 out of 3,360 galaxies, respectively. The cyan contours are rejuvenated objects with light-averaged ages smaller than $2.5\;$Gyr (see Fig.~\ref{fig:agestat}), the orange contours are the old red sequence population. Underlying grey contours include both populations. The solid line is a linear fit to the red population, the parameters of the fit are given at the top of each panel. The dotted line is the fit all environmental densities (parameters from top left-hand panel). Its distribution is shown by the top right-hand panel (same colour coding). The label gives the standard deviation $\sigma$ for the fit. The \aFe-$\sigma$ relationship for the red sequence population is independent of environment.}
\label{fig:afestat}
\end{figure*}
\begin{figure*}
\includegraphics[width=0.49\textwidth]{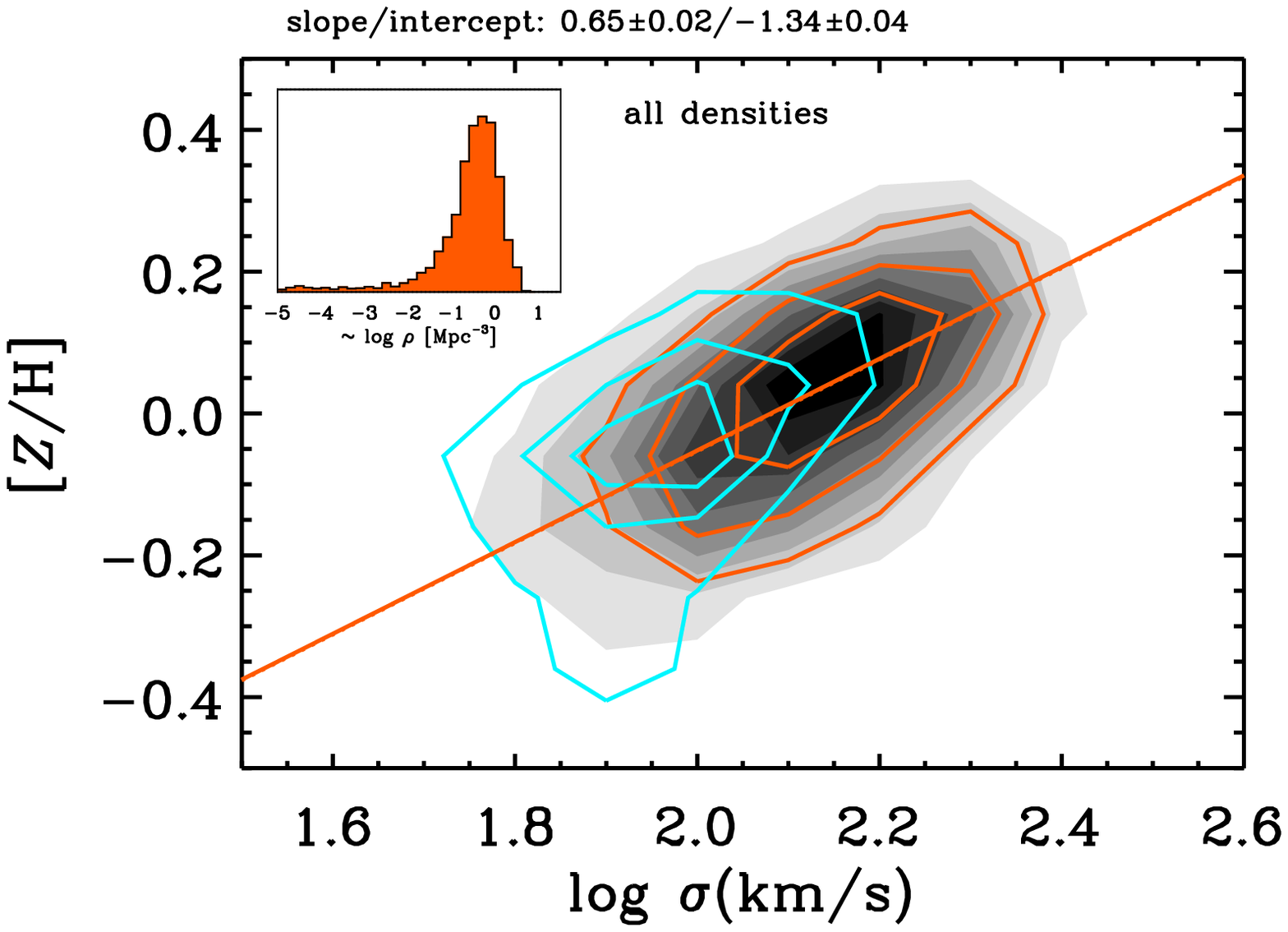}
\includegraphics[width=0.49\textwidth]{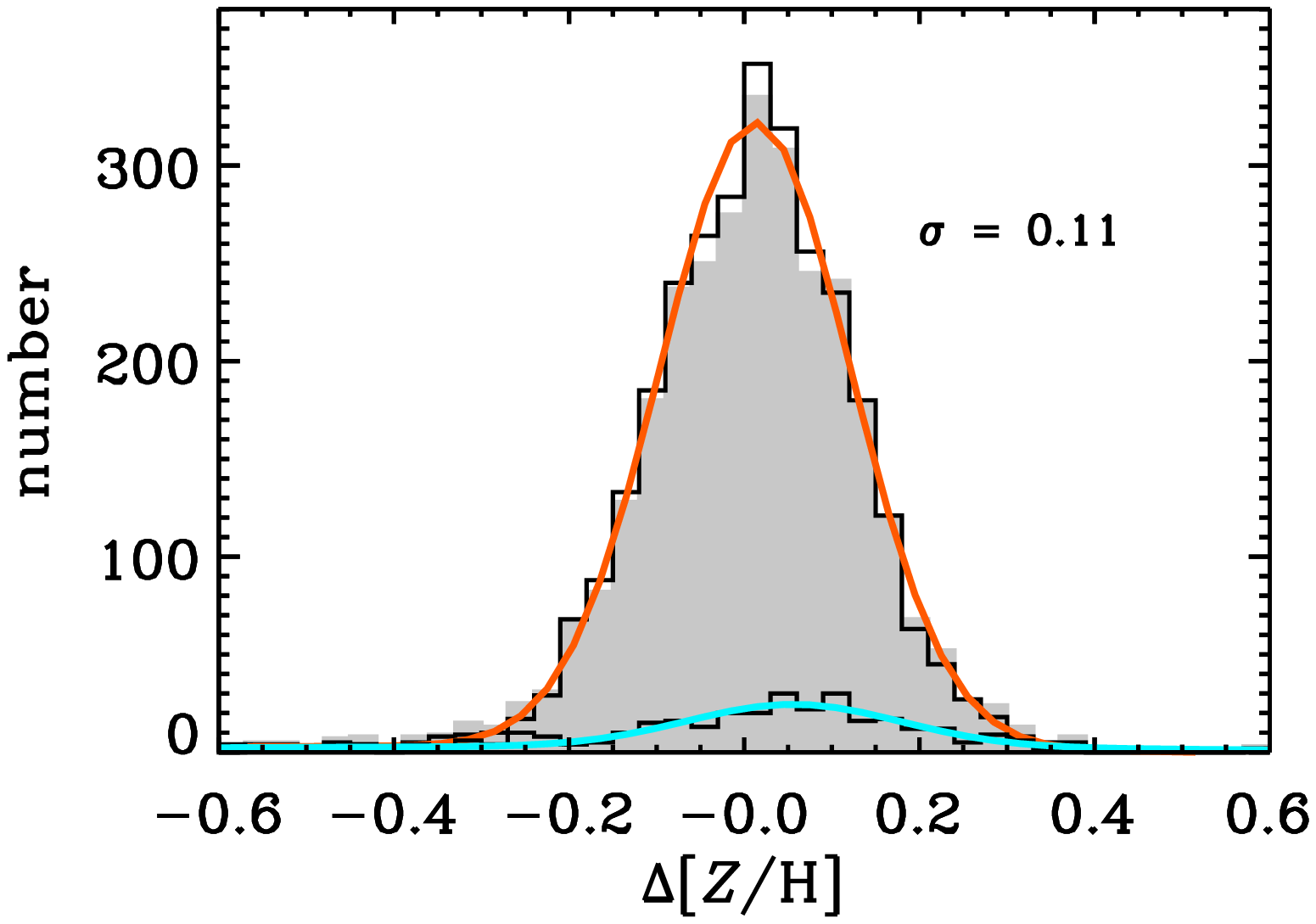}
\includegraphics[width=0.33\textwidth]{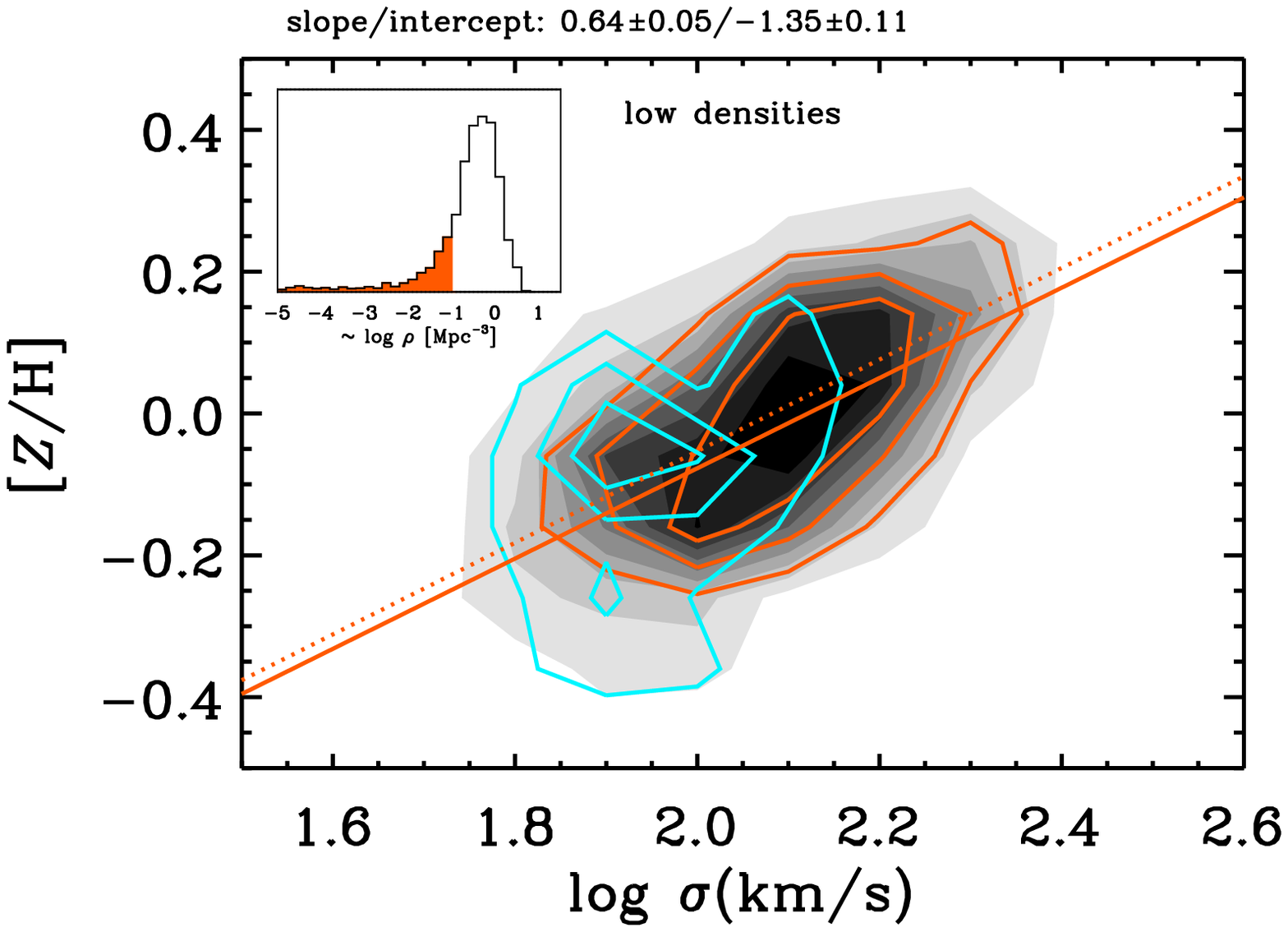}
\includegraphics[width=0.33\textwidth]{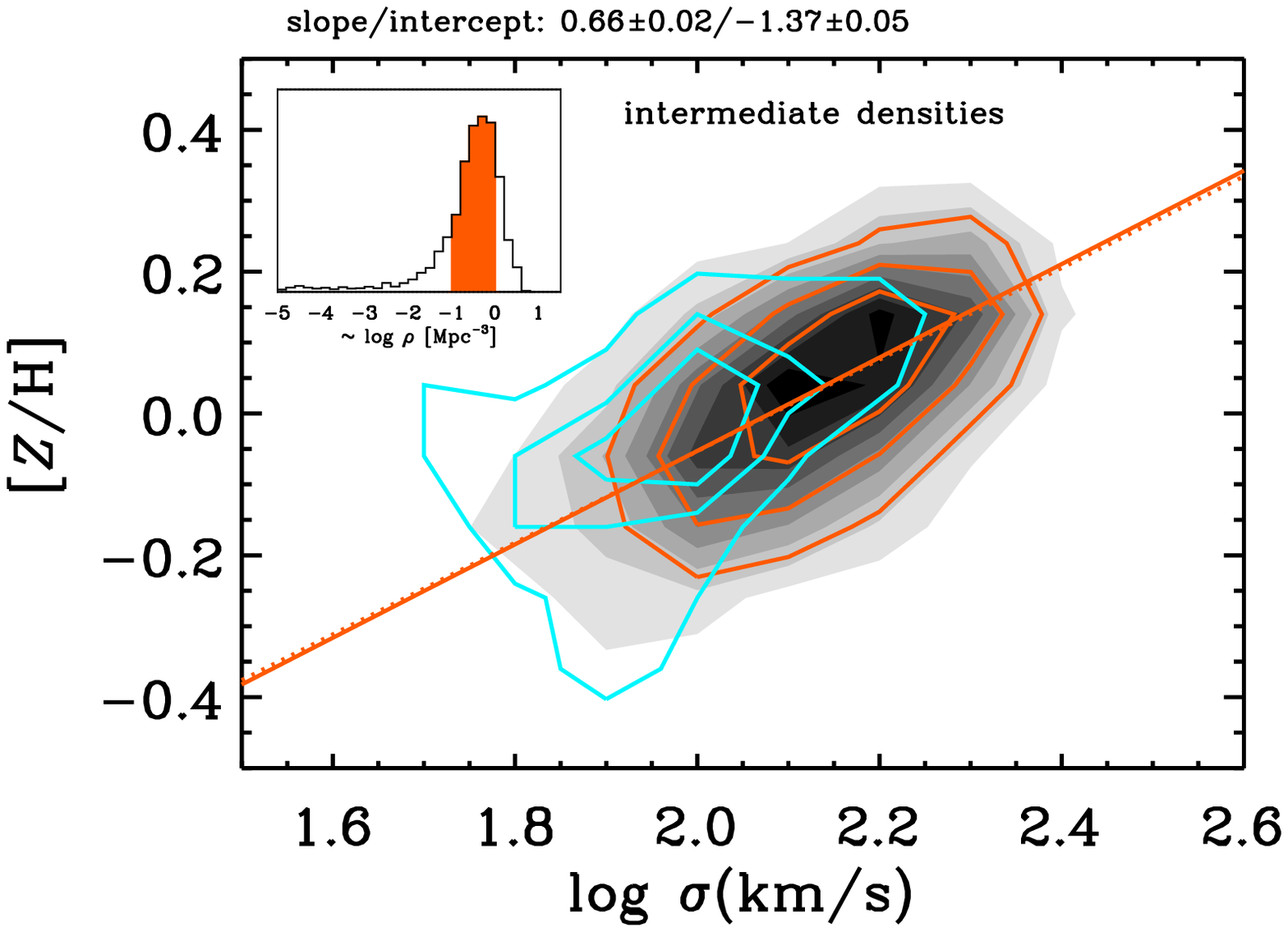}
\includegraphics[width=0.33\textwidth]{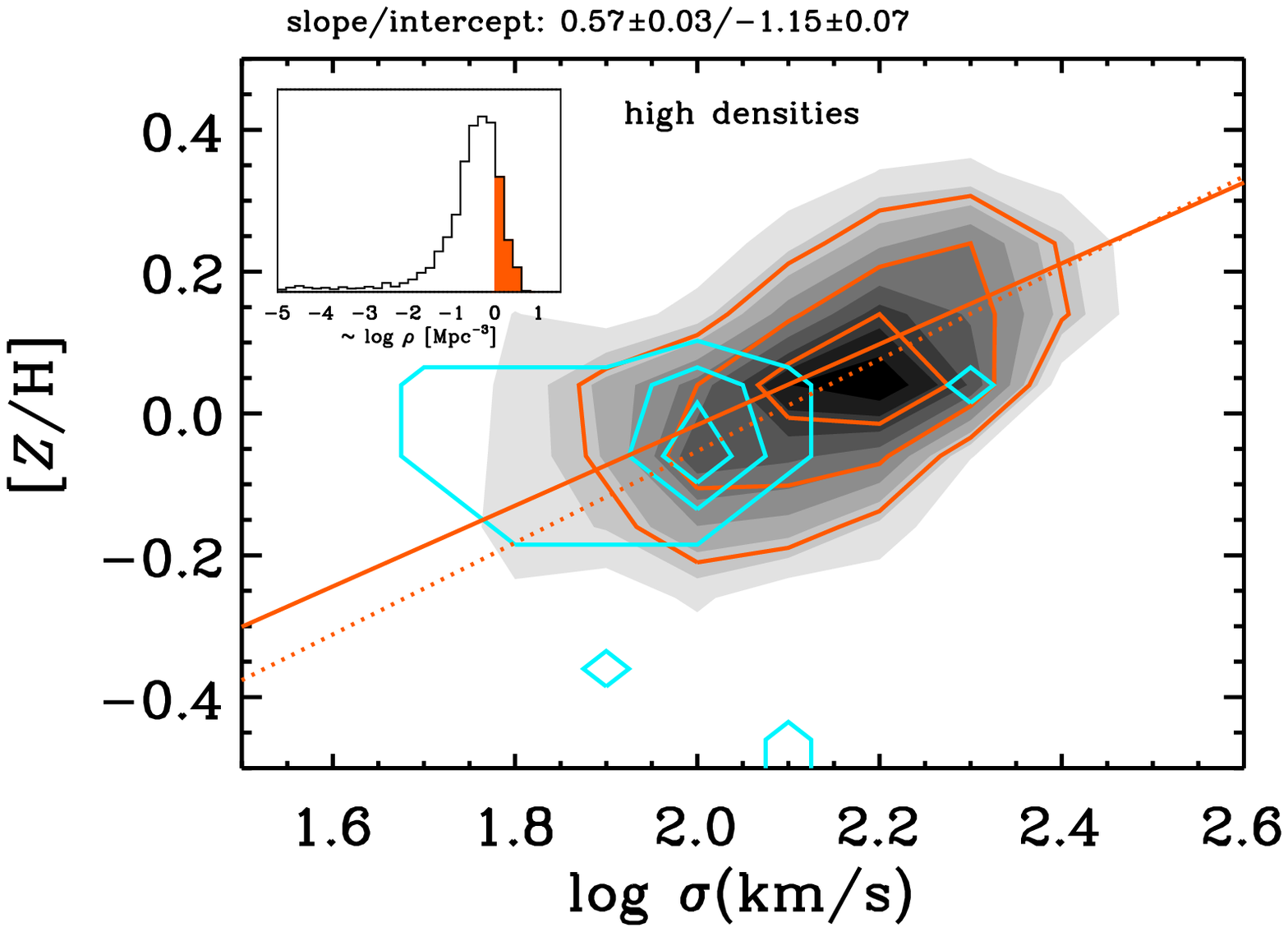}
\caption{Contour plots of the relationship between stellar velocity dispersion and total metallicity for various environmental densities as indicated by the inset histograms. The environmental density is proportional to the number density per volume, but no precise physical units are associated to it. The lowest and highest density bins contain 658 and 571 out of 3,360 galaxies, respectively. The cyan contours are rejuvenated objects with light-averaged ages smaller than $2.5\;$Gyr (see Fig.~\ref{fig:agestat}), the orange contours are the old red sequence population. Underlying grey contours include both populations. The solid line is a linear fit to the red population, the parameters of the fit are given at the top of each panel. The dotted line is the fit for all environmental densities (parameters from top left-hand panel). Its distribution is shown by the top right-hand panel (same colour coding). The label gives the standard deviation $\sigma$ for the fit. The \ZH-$\sigma$ relationship for the red sequence population is independent of environment.}
\label{fig:zhstat}
\end{figure*}

\subsection{Age as a function of environment}
Fig.~\ref{fig:agestat} shows contour plots for the relationship between stellar velocity dispersion and
luminosity-weighted age for various environmental densities as indicated by the inset histograms. The top left-hand panel includes all environments. The three bottom panels show three different choices of environments (orange histograms), i.e. the median density around the peak of the distribution (middle), and the lowest (left-hand panel) and highest (right-hand panel) density ends of the distribution. The latter are expected to show truly isolated and densest cluster galaxies and contain 658 and 571 out of 3,360 galaxies, respectively. Note that our density estimator is proportional to the number density per volume but does not have a strictly physical meaning, hence we do not attempt a more accurate identification with astrophysical objects like galaxy groups and clusters.

The top-left and right-hand panels show that the distribution of ages is bimodal with a major peak at old ages (orange contours) and a secondary peak at young ages around $\sim 2\;$Gyr (cyan contours) in analogy to 'red sequence' and 'blue' cloud identified in galaxy populations (see also Fig.~\ref{fig:sigmaur}). The sub-population of young early-type galaxies could be identified here mostly thanks to the purely morphological selection algorithm \citep[see also][]{Schawetal09b}. The dashed line at $\log\ ({\rm age})=0.4$ divides the two populations. The objects below this line have very young light-averaged ages ($<2.5\;$Gyr), and have most likely been rejuvenated by recent minor star formation events at lookback times $<2.5\;$Gyr. Note that a minor rejuvenation event adding only a few per cent of young stellar populations has a large effect on the luminosity-average age \citep[e.g.,][]{Greggio97,ST07}. Therefore, we call the objects below the dashed line 'rejuvenated', and the fraction of objects in this category 'rejuvenation fraction'. The latter is about 10 per cent as indicated by the label in the figure. It can also be seen that rejuvenation occurs mostly in lower-mass galaxies with $\log\sigma<2.2$ as previously found in other studies \citep[T05,][]{Yietal05,TD06,Donasetal07}.

We discuss the rejuvenated population in the 'blue cloud' separately and derive stellar population scaling relations for the 'red sequence' population that dominates the mass budget of the galaxy population. The linear fit to this population is shown by the solid line. The parameters of the fit are given at the top of the panel. The distribution about this fit is shown in the top right-hand panel (black histogram and orange line). The cyan line is the rejuvenated population and is clearly offset. 

The fits to the red sequence population are repeated in every environment bin. The dotted lines in all panels indicate the fit including all environmental densities. It can be seen clearly from the plots that the resulting fit parameters are consistent with no variation as a function of environment within their 1-$\sigma$ error bars. We conclude there is no considerable change as a function of environmental density, hence the age-$\sigma$ relationship is independent of environment.

Very different is the behaviour of the blue cloud (rejuvenated) population. In this case, the environment plays a
critical role. The rejuvenation fraction increases with decreasing environmental density. Hence early-type galaxies in lower density environments are not generally
younger (at a given mass), but the fraction of rejuvenated galaxies
is higher. This dependence on environment is
very similar to what has been found by \citet{Schawetal07a} albeit tracing
only the most recent residual star formation episodes in early-type galaxies
through near-$UV$ colours.

Blue horizontal branches from metal-poor or peculiar metal-rich
sub-populations \citep{GR90,DOR95,YDO98} may mimic an apparent presence of young populations \citep[][T05]{MT00,Leeetal00,Maretal03,Traetal05}. This degeneracy is very difficult to disentangle in unresolved stellar populations. We use therefore further indicators of residual star formation that are independent of horizontal branch morphology for consistency checks. The \aFe\ element abundance ratio is a useful quantity, because it quantifies the relative importance of delayed chemical enrichment from Type~Ia supernovae owing to late residual star formation \citep[e.g.,][]{GR83,MG86,Ma94,TGB99,Greggio05}. We further study emission line properties as indicators for possible star formation and/or AGN activity, as the ancillary use of emission line diagnostics allows us to break this degeneracy.

\subsection{\boldmath The \aFe\ ratio as a function of environment}
Fig.~\ref{fig:afestat} shows contour plots for the \aFe-$\sigma$ relations using the same
symbols and colour coding as in Fig.~\ref{fig:agestat}. We confirm
previous findings of a very well-defined and tight correlation between
velocity dispersion and \aFe\ ratio. The linear fits are again performed on the red sequence population (orange contours) as defined in the previous section. The distribution about this fit is shown in the top right-hand panel (black histogram and orange line). The cyan line is the rejuvenated population. The dotted lines in all panels indicate the fit including all environmental densities. Also in this case the slope and zero-point of the correlation are consistent with no variation as a function of environment.

The environment-dependent rejuvenated population (cyan contours) is offset from the general fit to slightly lower [\aFe] ratios by $\sim 0.1\;$dex. Note that errors in age and \aFe\ do not correlate such that they could create the offset in \aFe\ seen here (see Fig. 3 in Thomas et al 2005). This result is critical, as it supports the interpretation that the low light-averaged ages of the rejuvenated population are indeed caused by the presence of residual star
formation in these objects. This conclusion is further supported by the fact that rejuvenated elliptical galaxies with low \aFe\ ratios tend to have extra light in the centre pointing toward a dissipational formation scenario \citep{Kormendyetal09}. Fig.~\ref{fig:afestat} shows that the offset in \aFe\ ratio is relatively small, however, which suggests that only minor star formation episodes are
responsible for this rejuvenation. This is in agreement with the mean stellar ages derived from the mid-infrared fluxes of early-type galaxies that only allow for small amounts of recent star formation \citep{Temietal05}.

\subsection{Total metallicity as a function of environment}
The contour plots for the \ZH-$\sigma$ relations are shown in Fig.~\ref{fig:zhstat} using the same symbols and colour coding as in Fig.~\ref{fig:agestat}. Again, we confirm
the existence of a well-defined \ZH-$\sigma$ relation. Our data mainly improve on
its tightness and significance. The linear fits are again performed on the red sequence population (orange contours) as defined in the previous section. The distribution about this fit is shown in the top right-hand panel (black histogram and orange line). The cyan line is the rejuvenated population. The dotted lines in all panels indicate the fit including all environmental densities. Like for age and \aFe\ ratio, these relationships are independent of environment.

Fig.~\ref{fig:zhstat} further shows that the rejuvenated galaxies (cyan symbols) have slightly higher metallicities than the bulk population by $\sim 0.1\;$dex. This agrees well with previous findings of an age-metallicity anti-correlation at given $\sigma$ in early-type galaxies \citep[e.g.,][]{Traetal00b}. The slight increase in
metallically suggests that the residual star formation in these galaxies does not involve purely metal-poor pristine gas but must include at least some fraction of previously enriched interstellar
medium, either from internal re-processing or external accretion. Again, the fact that the offset is small suggests that the rejuvenation event involves only a minor fraction in mass.

\subsection{Colour}
\begin{figure}
\includegraphics[width=0.49\textwidth]{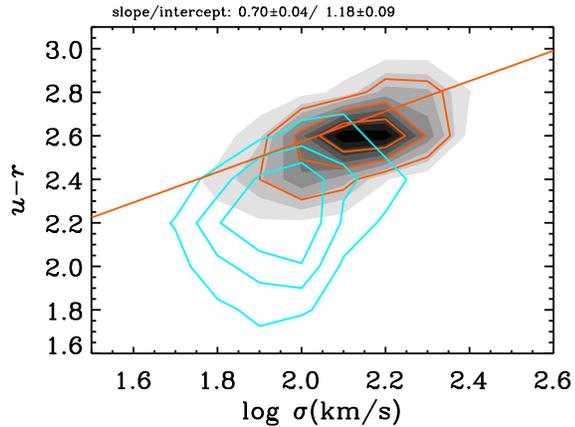}
\caption{Contour plots of the relationship between stellar velocity dispersion and \ur\ colour including all environmental densities. The cyan contours are rejuvenated objects with light-averaged ages smaller than $2.5\;$Gyr (see Fig.~\ref{fig:agestat}), the orange contours are the old red sequence population. Underlying grey contours include both populations. The solid line is a linear fit to the red population, the parameters of the fit are given at the top of the panel. The rejuvenated population is clearly visible through its blue \ur\ colour, which is offset from the general trend by $\sim 0.2\;$mag.}
\label{fig:sigmaur}
\end{figure}
As a cross-check of the stellar population parameters derived from absorption line indices we briefly discuss \ur\ colour. Fig.~\ref{fig:sigmaur} shows contour plots for the \ur-$\sigma$ relation using the same
symbols and colour coding as in Fig.~\ref{fig:agestat}. All environmental densities are included.
The linear fits are again performed on the red sequence population (orange contours) as defined in the previous section. The rejuvenated population (cyan contours) with light-averaged ages below $2.5\;$Gyr as defined in Fg.~\ref{fig:agestat} is well separated from the red population showing bluer \ur\ colour by about $0.2\;$mag. The old population forms a well-defined red sequence, while the rejuvenated objects occupy the region that is known as the 'blue cloud'. The \ur\ colour clearly reinforces the conclusions drawn from Figs.~\ref{fig:agestat}.

\subsection{Star formation activity and AGN}
It is interesting to analyse the ionisation states of the gas in the early-type galaxies as a function of their position on the age-$\sigma$ relationship. In particular, the possibility of blue horizontal branch stars mimicking the presence of young stellar populations can be ruled out if rejuvenation coincides with star formation activity detected through emission lines. The majority of the sample (77 per cent) shows no significant emission lines, hence are classified as 'passive'. To the remaining 'active' 23 per cent we apply the BPT classification scheme that separates star formation from AGN \citep{1981PASP...93....5B,1987ApJS...63..295V,2001ApJ...556..121K,2003MNRAS.346.1055K,2003ApJ...597..142M}. A detailed BPT analysis of the present sample is presented in \citet{Schawetal07b} and we refer the reader to this paper for further details. Using the classifications from \citet{Schawetal07b}, we find that 16 per cent of the objects with emission lines are star forming, 21 per cent are star forming/AGN composites, 16 per cent have AGN Seyfert-like emission lines, and 47 per cent show LINER-type emission.

\begin{figure}
\includegraphics[width=0.49\textwidth]{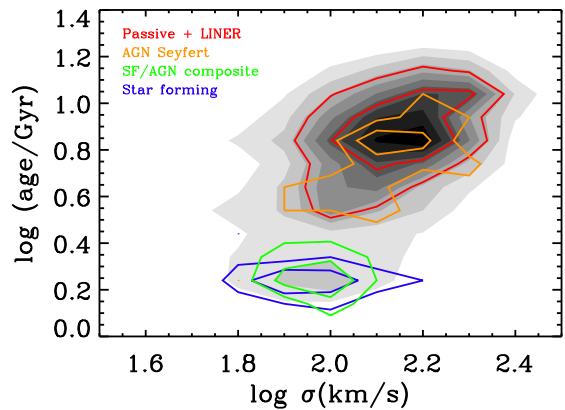}
\caption{Contour plots of the relationship between stellar velocity dispersion and age including all environmental densities (analogue to Fig.~\ref{fig:agestat}). The coloured contours are sub-populations separated through their emission line classifications star forming (blue), star forming/AGN composite (green), AGN Seyfert-like emission (orange), AGN LINER-like emission and passive (red). The underlying grey contours include all galaxies. All objects with traces of ongoing star formation (blue and green contours) belong to the rejuvenated population defined in Fig.~\ref{fig:agestat}, while galaxies with AGN have considerably older stellar populations.}
\label{fig:sigmaagebpt}
\end{figure}

In Fig.~\ref{fig:sigmaagebpt} we plot contours for these sub-populations over the age-$\sigma$ relationship of the whole sample (analogue to Fig.~\ref{fig:agestat}).  Star forming objects are plotted in blue, star forming/AGN composite objects in green, AGN with Seyfert-like emission in orange, and finally AGN with LINER-like emission and passive galaxies are plotted in red contours. It turns out that all objects that show signs of star formation activity (blue and green contours) have rejuvenated stellar populations. Some fraction of these are star forming/AGN composites, but all objects that host AGN in their centres without any star formation harbour old stellar populations.
More specifically, we have detected star formation activity in 286 out of 3360 objects (8.51 per cent) compared to 341 (10.15 per cent) that have been classified as rejuvenated. This implies that the majority of the rejuvenated galaxies (84 per cent) are actively forming stars, which is well consistent with the detection of significant amounts of neutral hydrogen in low-mass ellipticals \citep{Morgantietal06}. This clearly rules out the possibility of blue horizontal branch stars mimicking the presence of young stellar populations.

\subsection{Dynamical masses}
\begin{figure*}
\includegraphics[width=0.33\textwidth]{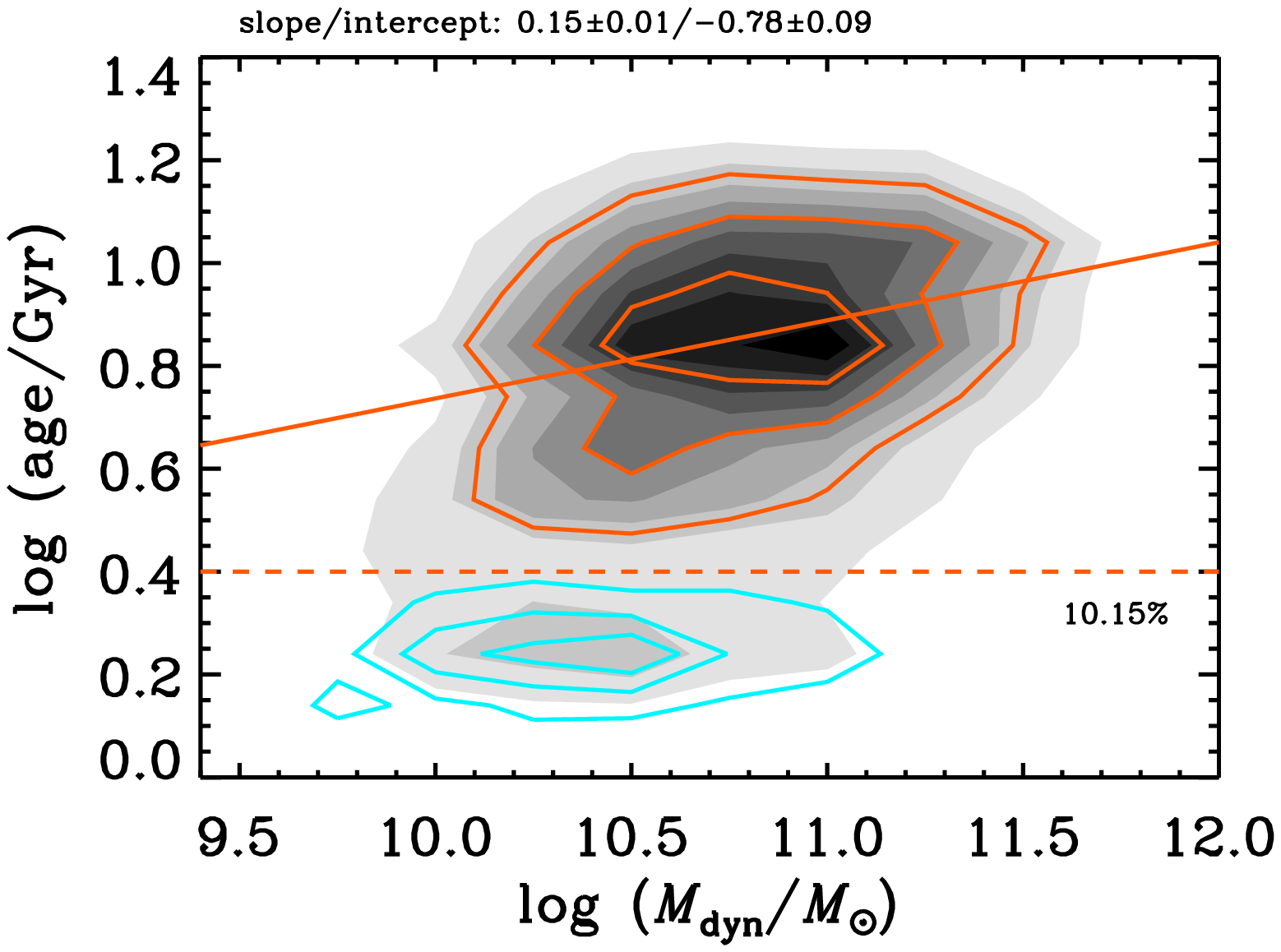}
\includegraphics[width=0.33\textwidth]{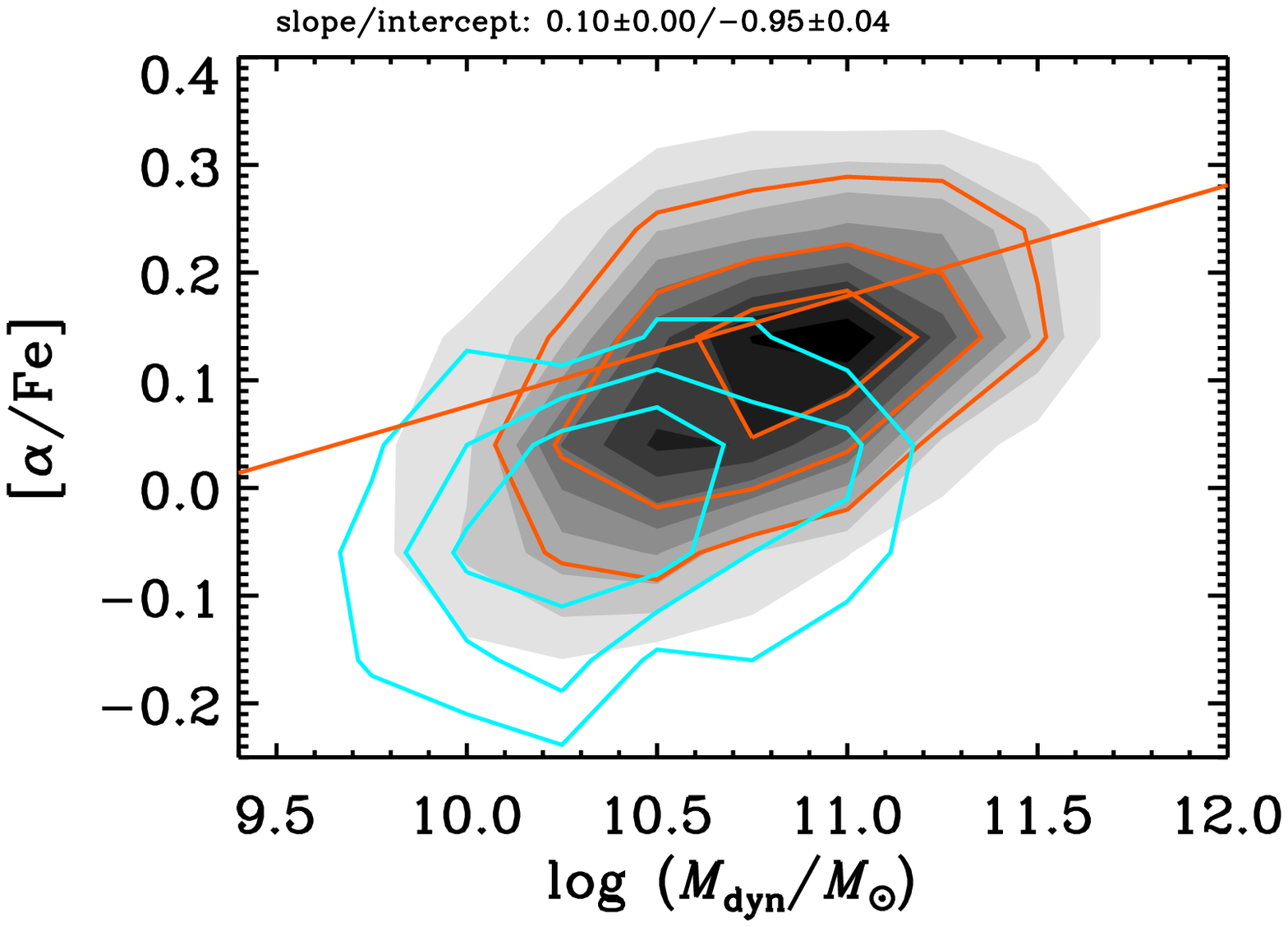}
\includegraphics[width=0.33\textwidth]{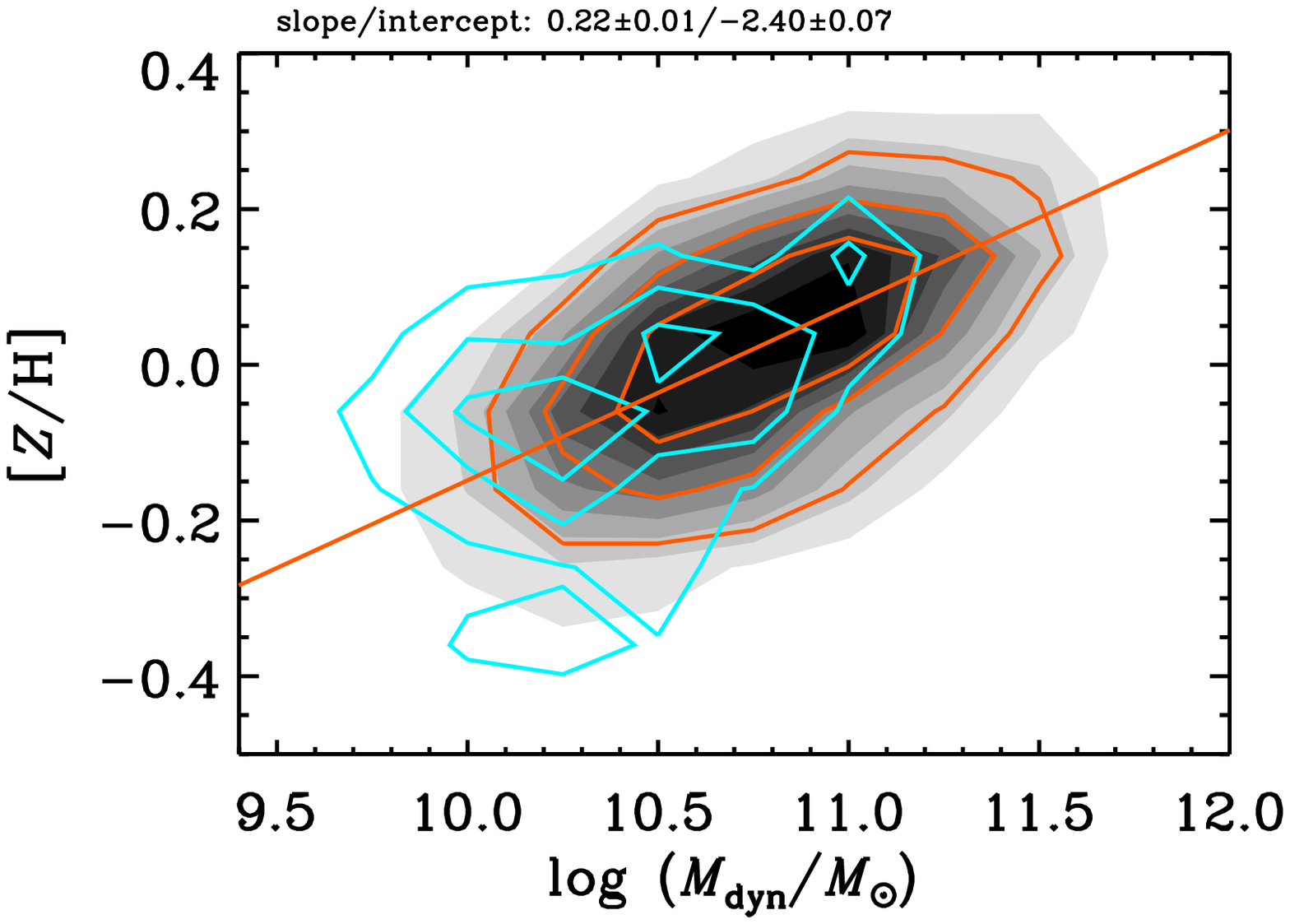}
\caption{Contour plots of the relationship between dynamical mass and light-averaged age (left-hand panel), \aFe\ element ratio (middle panel), and total metallicity (right-hand panel) including all environmental densities. The dashed line (left-hand panel) separates the old red sequence population (orange contours) from the rejuvenated objects with light-averaged ages smaller than $2.5\;$Gyr (cyan contours). The fraction of this latter population is given by the label. The underlying grey contours include both populations. The solid line is a linear fit to the red population, the parameters of the fit are given at the top of each panel. The dynamical mass is calculated from the scaling with velocity dispersion and effective radius adopted from \citet{Capetal06}.}
\label{fig:massplot}
\end{figure*}
So far in this paper we have used line-of-sight stellar velocity dispersion $\sigma$ as a proxy for galaxy mass. The major reason is that $\sigma$ is independent of the stellar population properties derived here and can be easily measured in galaxy spectra from the broadening of absorption line features \citep[e.g.,][]{FJ76,Bender90a}. However, $\sigma$ is a difficult quantity to predict in galaxy formation models, the direct output of which are stellar and dynamical masses. Hence, we determine dynamical masses from 
the scaling between virial dynamical galaxy mass $M_{\rm dyn}$ with stellar velocity dispersion $\sigma$, and effective radius provided by \citet{Capetal06}. Effective radii are taken from the SDSS database. The typical error in $\log\ M_{\rm dyn}$ is $0.05\;$dex with a tail extending to $\sim 0.2\;$dex. Note that this 'dynamical' mass should be a good approximation also of the baryonic mass, as baryonic matter is dominating the mass budget in early-type galaxies inside the effective radius \citep{Capetal06,Thoetal07}. This is different from the approach in T05, where a scaling with observed luminosity through model $M/L$ ratios as a function of age and metallicity is used. We prefer the dynamical mass estimates, as they are independent of the stellar population parameters derived.

Fig.~\ref{fig:massplot} presents the equivalent plots to Figs.~\ref{fig:agestat}, \ref{fig:afestat}, and \ref{fig:zhstat} showing the resulting relationships between dynamical mass and stellar population parameters light-averaged age (left-hand panel), \aFe\ element ratio (middle panel), and total metallicity (right-hand panel). All environmental densities are included. The dashed line (left-hand panel only) separates again the red sequence population (orange contours) from the rejuvenated objects with light-averaged ages smaller than $2.5\;$Gyr (cyan contours), while the underlying grey contours include both populations. The solid line is a linear fit to the red sequence population, the parameters of this fit are given at the top of each panel. It is evident that the scaling relations are just as well defined for dynamical mass as they are for velocity dispersion.

\subsection{Rejuvenation fractions}
The results presented in this work show that it is the fraction of rejuvenated galaxies (galaxies in the blue cloud) that is dependent of environment, while the stellar population scaling relations of the red sequence galaxies are environment independent. So far, we have not disentangled this from the general correlation between age and mass. As the rejuvenation fraction increases also with decreasing velocity dispersion and galaxy mass, its environmental dependence may just be caused by mass segregation, which is the fact that more massive galaxies live preferentially in denser environments.

\begin{figure*}
\includegraphics[width=0.8\textwidth]{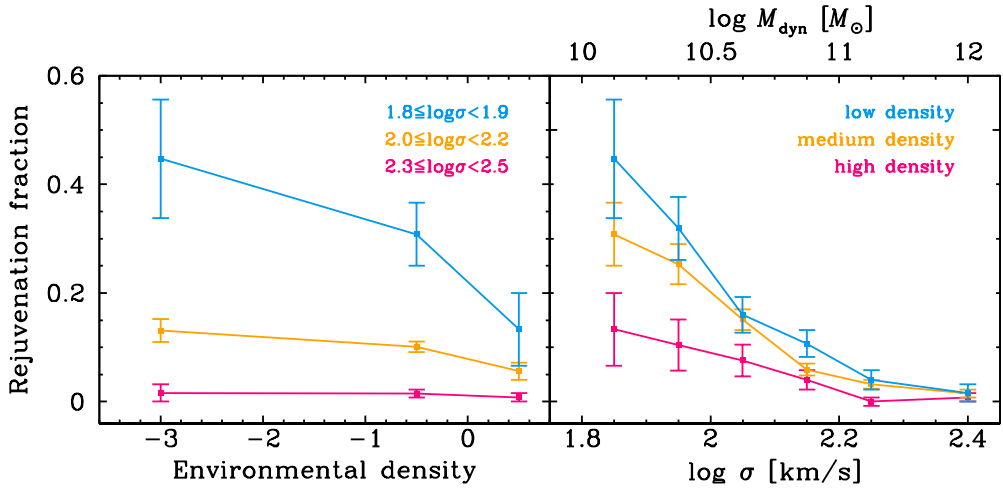}
\caption{{\em Left-hand panel:} Fraction of rejuvenated galaxies as a function of environmental
density (left-hand panel) for three bins in velocity dispersion as indicated by the labels. Note that
the units of environmental density provide a relative measure of local density and are not associated to astrophysical objects such as field, groups or clusters. {\rm Right-hand panel:} Fraction of rejuvenated galaxies as a function of galaxy velocity dispersion and dynamical mass (see top x-axis) for three different environmental density intervals $-5\leq \log\rho<-1$ (blue line), $-1\leq \log\rho<0$ (orange line), and $0\leq \log\rho<1$ (magenta line). Error bars are Poisson errors. The figure shows clearly that the rejuvenation fraction depends on both galaxy mass and environmental density.}
\label{fig:rsf}
\end{figure*}
Fig.~\ref{fig:rsf} quantifies this in more detail. The plots show the
fraction of rejuvenated galaxies as a function of environmental
density (left-hand panel) and velocity dispersion (or galaxy mass, right-hand panel) for three different bins in velocity dispersion and environmental density, respectively. Fig.~\ref{fig:rsf} demonstrates that the fraction of rejuvenated early-type galaxies increases as a function of both decreasing galaxy mass \citep[see also][]{Schawetal06} and decreasing environmental density \citep{Schawetal07a}, with the effect of environment being slightly weaker. Fractions vary from only a few per cent for the most massive galaxies in densest regions up to 45 per cent for the smallest ellipticals (a few $10^{10}\;$\Msun) in the least dense regions of the universe. It can be seen that the rejuvenation fractions as well as their dependence on environment are largest for the smallest galaxies (blue line in the left-hand panel). The most massive early-type galaxies have experienced no rejuvenation event independent of environment (magenta line in the left-hand panel). The rejuvenation fraction among low- and intermediate mass galaxies (blue and orange lines) increases from the highest to the lowest density bin at $>2\sigma$ significance.
The increase of the rejuvenation fraction as a function of velocity dispersion is most pronounced for the lowest environmental densities (blue line in right-hand panel).

\subsection{The final scaling relations}
The age fits for the red sequence population shown in Figs.~\ref{fig:agestat} and \ref{fig:massplot} as orange solid lines are $\log\ {\rm age}({\rm Gyr}) = -(0.25\pm 0.05) + (0.52\pm 0.02)\ \log\sigma$ and $\log\ {\rm age}({\rm Gyr}) = -(0.76\pm 0.09) + (0.15\pm 0.01)\ \log M_{\rm dyn}$.  Note, however, that the sample discussed here covers the
redshift range $0.05\leq z\leq 0.06$. Even though these are very
moderate redshifts, they correspond to look-back times that are
significant for the purposes of this work \citep[see also][]{Beretal06}.  Adopting $\Omega_m=0.24$,
$\Omega_\Lambda=0.76$, and $H_0=73\;$km/s/Mpc
\citep{Tegmarketal06,Percetal07} we obtain an average lookback time of $700\;$Myr for this redshift interval. By correcting for this time
evolution (on the linear age scale), the zero points of the above
relationships (on logarithmic scale) increase to $-0.11$ and $-0.53$, and the slopes
decrease to 0.47 and 0.13. We assume that both metallicity and \aFe\ ratio
do not evolve.

We then obtain the following {\em local} stellar population scaling
relations, age (log ($t$/Gyr)), total metallicity (\ZH), and \aFe\ element
ratio as functions of velocity dispersion $\sigma$ (in km/s) and dynamical mass $M_{\rm dyn}$ (in $M_{\odot}$). Note that these are {\em independent of the environment}.
\begin{eqnarray}
\label{eqn:relations}
\log\ t &=& -(0.11\pm 0.05) + (0.47\pm 0.02)\ \log (\sigma)\\\nonumber
[\ZH] &=& -(1.34\pm 0.04) + (0.65\pm 0.02)\ \log (\sigma)\\\nonumber
[\aFe] &=& -(0.55\pm 0.02) + (0.33\pm 0.01)\ \log (\sigma)
\end{eqnarray}
\begin{eqnarray}
\label{eqn:Mrelations}
\log\ t &=& -(0.53\pm 0.09) + (0.13\pm 0.01)\ \log (M)\\\nonumber
[\ZH] &=& -(2.40\pm 0.07) + (0.22\pm 0.01)\ \log (M)\\\nonumber
[\aFe] &=& -(0.95\pm 0.04) + (0.10\pm 0.01)\ \log (M)
\end{eqnarray}

These relationships represent the red sequence population and are valid for a large range in galaxy masses ($\sigma\ga 100\;$kms$^{-1}$ or $M_{\rm dyn}=3\cdot 10^{10}\;\Msun$) and all environmental densities. Note that a major fraction of galaxies deviates from these scaling relations at the low ends of the mass and environmental density distributions, as the fraction of rejuvenated galaxies increases to $\sim 45\;$ per cent (see Fig.~\ref{fig:rsf}).

The observed spread in log age, metallicity, and [\aFe] ratio in
Eqn.~\ref{eqn:relations} is on average $0.23\;$dex, $0.11\;$dex, and
$0.07\;$dex, respectively (see top right-hand panels in Figs.~\ref{fig:agestat}, \ref{fig:afestat}, and \ref{fig:zhstat}). Subtracting the errors estimated in
Section~\ref{sec:data}, we obtain an average {\em intrinsic} scatter
of $0.21\;$dex, $0.08\;$dex, and $0.02\;$dex, respectively. The intrinsic
scatter in age increases slightly to $0.27\;$dex at the lowest environmental densities, while the intrinsic spread in both metallicity and \aFe\ is virtually independent of the environment. 

The slopes and zero-points for the scaling relations found in this work are in reasonable agreement with other recent results in the literature (see references in the Introduction). The slope of the age-$\sigma$ relation is somewhat steeper than what we have found in T05. In particular the results for the high-density sample in T05 are different, which might well be caused by a particularity of the Coma Cluster that dominates the high-density part of the sample in T05. Still, the impact on the derived formation epochs and their dependence on galaxy mass is minor (see Section~\ref{sec:epochs}). The most important difference with T05 and other previous work is that the stellar population scaling relations are independent of the environment for the bulk of the population.

\section{Epochs of early-type galaxy formation}
\label{sec:epochs}
The luminosity-weighted ages and \aFe\ element ratios derived here
enable us to approximate star formation histories. This is possible as
the element ratio constrains formation time-scales. By means of a
chemical evolution model \citep{TGB99}, in T05 we determined a simple scaling between \aFe\ ratio and formation time-scales. The latter together with the average age defines a star formation history. The resulting star formation rates as functions of lookback time for various mass bins are shown in Fig.~\ref{fig:sfr}. Masses are derived from velocity dispersion and effective radius through the scaling provided by \citet{Capetal06}. As discussed in detail in T05, these star formation histories are meant to sketch the typical formation history averaged over the entire early-type galaxy population (at a given mass). Real star formation histories of individual objects are expected to be more bursty and irregular.

\subsection{Comparison with T05}
The formation epochs of early-type galaxies below $\sim 10^{11}\;\Msun$ are in good agreement with the values derived in T05 for the low-density sample. At the highest masses the situation is more complex. The steeper slope of the age-$\sigma$ relationship found here with respect to the low-density sample of T05 leads to significantly larger formation ages of the most massive early-type galaxies with $\log\sigma\sim 2.5$ ($\sim 10^{12}\;\Msun$) by about $2\;$Gyr. The lookback time for their formation is $\sim 12\;$Gyr (instead of $\sim 9.5\;$Gyr in T05) and hence conspires to be very similar to the ages of the high-density massive early-types in T05. Their formation redshifts, however, are lower ($\sim 4$) than in T05 ($\sim 5$) because of the lower values for both matter density and Hubble parameter adopted in the present work following the most recent determinations \citep{Tegmarketal06,Percetal07}.

\subsection{Archaeological downsizing}

\begin{figure*}
\centering\includegraphics[width=0.8\textwidth]{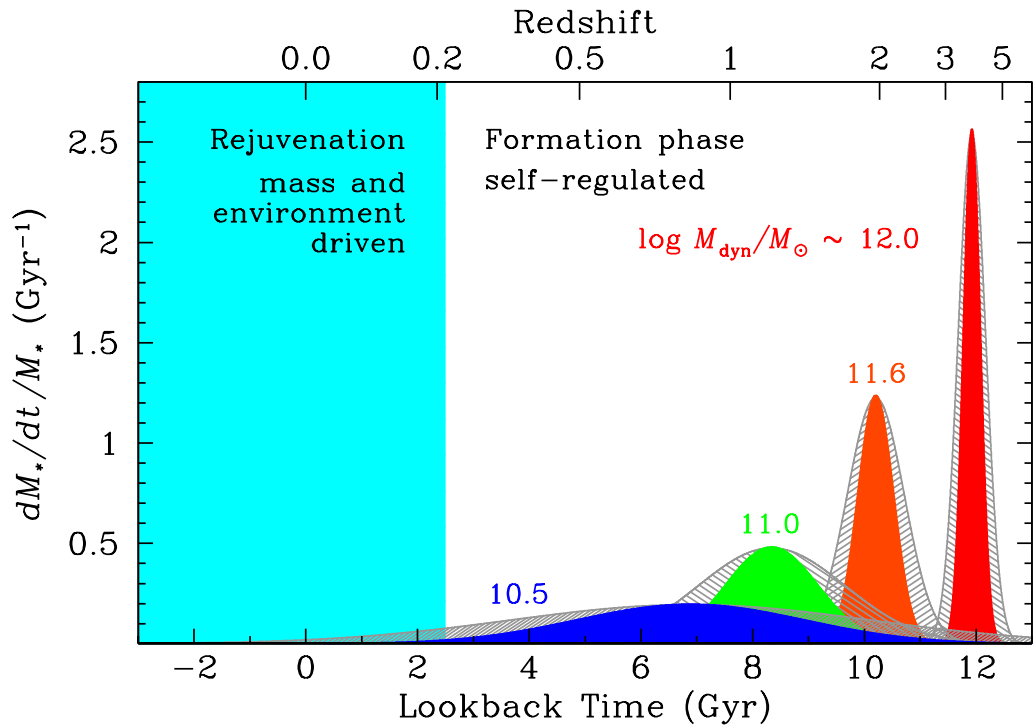}
\caption{Specific star formation rate as function of look-back time
for early-type galaxies of various masses as indicated by the labels. The grey hatched curves
indicate the range of possible variation in the formation time-scales
that are allowed within the {\em intrinsic} scatter of the \aFe\
ratios derived. No dependence on environmental density is found. The upper x-axis connects time and redshift adopting $\Omega_m=0.24$,
$\Omega_\Lambda=0.76$, and $H_0=73\;$km/s/Mpc. Note that these star formation histories are meant to sketch the typical formation history averaged over the entire galaxy population (at a given mass). Real star formation histories of individual objects are expected to be more bursty and irregular. Intermediate- and low-mass galaxies in low-density environments get rejuvenated via minor star formation events below redshift $z\sim 0.2$ (see Section 3.8). This suggests a phase transition from a self-regulated formation phase without environmental dependence to a rejuvenation phase, in which the environment plays a decisive role possibly through galaxy mergers and interactions.}
\label{fig:sfr}
\end{figure*}
Fig.~\ref{fig:sfr} illustrates the key observation that more massive
galaxies form at higher redshift, now commonly referred to as
'downsizing'.  This naming was first suggested by \citet{Cowetal96}
who studied deep survey data and found that 'the maximum rest-frame
$K$ luminosity of galaxies undergoing rapid star formation has been
declining smoothly with decreasing redshift'. The present approach
based on the 'archaeology of stellar populations', instead, manifests
downsizing for the local galaxy population, which logically is called
'archaeological downsizing' by \citet{NvdBD06}.  Note that the latter
is a universal characteristic of galaxy formation and is not
restricted to early-types \citep{Gavetal02,Heavensetal04,CidFeretal07,Panteretal07}.
Most importantly, the convergence of local and high redshift
observations to the general downsizing pattern is vital to set
stringent constraints on the theory of galaxy formation.

Clearly, a mechanism is needed to regulate star formation in massive
objects in order to produce the anti-hierarchical trend suggested by
observations. AGN feedback has been spotted as the most promising
solution \citep{SR98,Binney04,Graetal04,Silk05,Milleretal06,Schawetal06,Hopkinsetal08,Renzini09,2009Natur.460..213C,Kormendyetal09}. In a companion paper exploiting the MOSES sample \citep{Schawetal07b,Schawetal09a} we have found first empirical evidence that such a mechanism may indeed be occurring in early-type galaxies also at recent epochs. Moreover, AGN outflows at high redshift provide observational evidence for AGN feedback at early epochs in the evolution of galaxies \citep[e.g.][]{Nesvadba08}.
The most recent renditions of semi-analytic, hierarchical galaxy formation models include a prescription of AGN feedback \citep{DeLuciaetal06,Crotonetal06,Boweretal06,MFT07,Cattetal05,Cattetal06,DB06,2007MNRAS.377...63C,2007arXiv0704.3941C}. \citet{DeLuciaetal06}
show that the simulations now produce star formation histories
that are much closer to the observational constraint presented in T05
(and Fig.~\ref{fig:sfr} of the present work) than previous generations
of semi-analytic models. The observed \aFe\ enhancement of early-type galaxies and in
particular its correlation with galaxy mass (Fig.~\ref{fig:afestat}) needs yet to be reconciled with hierarchical models of galaxy formation \citep{Thomas99a,Nagaetal05,Pipetal09}. In \citet{Pipetal09} it is shown that the predicted \aFe-mass relationship is still flatter and has considerably more scatter than the observations even in models with AGN feedback. This conclusion is confirmed by the recent study of \citet{Arrigonietal09}, who show that the \aFe-$\sigma$ relation can only be produced if a top-heavy initial mass function and a lower fraction of binaries that explode as Type Ia supernovae are adopted.

\subsection{\boldmath Phase transitions in the local universe}
A fundamental difference with respect to previous results is the lack of a
dependence on environment. Fig.~10 in T05 presenting the formation epochs shows a significant delay in the formation of massive galaxies in the field. As discussed in T05, this suggested that massive
galaxies in low densities form later but on the same time-scales as
their counterparts in clusters (see Fig.~10 in T05) as the direct
consequence of early-type galaxies in denser environments having higher ages
but the same \aFe\ ratios. This implied that star formation must have
essentially been on hold for the first $3\;$Gyr in massive objects in
low densities. The new (and statistically more robust) results presented here suggest a different picture. In the new version of the star formation history plot (see Fig.~\ref{fig:sfr}) the main formation epochs of early-type galaxies as a function of galaxy mass are independent of the environmental density.

However, some early-type galaxies are rejuvenated, i.e.\ they must have harboured minor star formation events at recent epochs within the past few Gyrs (see Section~3). Such recent star formation activity that we call 'rejuvenation' occurs {\em on top} of the star formation histories shown in Fig.~\ref{fig:sfr}. The fraction of this rejuvenated galaxy population increases both with decreasing galaxy mass and with decreasing environmental density (see Fig.~\ref{fig:rsf}). This implies that the impact of environment increases with decreasing galaxy mass \citep{Tasca09}. The dependence on environment suggests that galaxy interactions and merger activity might have been the major triggers of early-type galaxy rejuvenation in the past few billion years since approximately redshift $z\sim 0.2$. During most of the Hubble time, instead, the formation and evolution of early-type galaxies was driven mainly by self-regulation processes and intrinsic galaxy properties such as mass. The stellar populations in early-type galaxies did not seem to feel in which environment the galaxy was forming. It is during this formation phase before $z\sim 0.2$ when the bulk of the stellar populations in early-type galaxies formed, and when the stellar population scaling relations (see Eqns.~\ref{eqn:relations} and~\ref{eqn:Mrelations}) were established. This stage is then followed by a rejuvenation phase at redshifts below $z\sim 0.2$ affecting mostly low- and intermediate mass early-type galaxies, during which the environment must have played a key role possibly via galaxy mergers and interactions.

Thus, there must have been a phase transition a few billion years ago in the evolution of early-type galaxies from an environment-independent self-regulated phase of formation to an environment-dependent phase of (possibly) merger-driven rejuvenation as sketched in Fig.~\ref{fig:sfr}. The universe may eventually move into a secular evolution phase in which intrinsic galaxy properties re-gain dominance over galaxy interactions as the main driver of galaxy evolution \citep{KK04}.

\section{Discussion}
\label{sec:discussion}
In this paper we investigate the effect of environmental density on the formation epochs of early-type galaxies. The major improvement with respect to T05 and other previous work in the literature is the significant increase in sample size allowing for a proper statistical analysis. We study 3,360 elliptical galaxies in a very narrow redshift bin ($0.05\leq z\leq 0.06$) that have been morphologically selected from the SDSS DR4. The most important result is that, at fixed galaxy mass, we find no dependence of the stellar population parameters on environmental density for the bulk of the population. It is the fraction of rejuvenated ellipticals (about 10 per cent in the average environment), instead, that increases strongly with decreasing environmental density. In the following we compare this finding to other recent work in the literature. 

As outlined in the Introduction, pre-SDSS studies of the stellar population
parameters in early-type galaxies based on relatively small, local samples consistently find younger average ages in low density environments. T05 have found an average shift in the luminosity-weighted age by $\sim 2\;$Gyr. The present sample based on SDSS is significantly larger and therefore less biased towards the local volume, which may explain the discrepancy with previous work. The increase in sample size further implies a higher statistical significance of the results.

A first step toward large samples beyond the very local universe has been done by \citet{Beretal98} who
analyse the \Mgtwo-$\sigma$ relationship of a sample of 931 early-type
galaxies in various environments. Interestingly, they find the
influence of the environment to be negligible, in agreement with the
present results. A more detailed comparison is not possible, however,
as \citet{Beretal98} do not derive stellar population parameters
directly. In a more recent study based on the SDSS, instead,
\citet{Beretal06} discuss the stellar population scaling relations as a
function of environment. Their Fig.~14 clearly agrees with our
conclusion that these relationships are insensitive to the
environmental density.

\citet{Smithetal06} analyse 3,000 red-sequence galaxies in nearby
clusters and find some variation of the stellar population parameters
along the cluster radius. Early-type galaxies appear about $0.06\;$dex
younger and slightly less \aFe\ enhanced in the outskirts. This result
is not necessarily inconsistent with the present study. This effect on the average quantities is small and might well be driven by the increase in the relative number of rejuvenated ellipticals rather than by a change in properties of the bulk population.

Interesting in the context of this paper is also the work by
\citet{Hainesetal06} based on SDSS DR4. In essence they find that the
typical ages of dwarf galaxies show strong dependence on environment, from being relatively old, passive satellites in clusters to being actively star forming and rejuvenated systems in the field.  This is
very well in line with our finding that it is the rejuvenation
fraction (which is largest in the smallest galaxies) that strongly
depends on environment.

Colour gradients (with bluer colours at larger radii) are found to
become steeper with decreasing environmental density
\citep{LaBarberaetal05,KI05}. Combined with the fact that colour
gradients are produced by metallicity rather than by age
\citep{Sagetal00,Tamuraetal00,Mehetal03,Wuetal05}, this suggests that
a higher (dry) merger rate in denser environments might have flattened
existing metallicity gradients. However, the results of the present
work imply that the formation and evolution of the stellar populations
in galaxies must still be controlled intrinsically independent of the
environment. Such dissipation-less, dry mergers in dense environments
have indeed been observed \citep{vDoketal99,Tranetal05}. But they must occur relatively late in the evolutionary history of the early-type
galaxy population and must be moderate in number in order to keep the stellar population scaling relations intact. The observationally constrained, modest merger rate for massive galaxies of one merger per galaxy since $z\sim 1$ may be consistent with that requirement \citep{Belletal06a,Pozzetti10}.

Most recently, \citet{2009arXiv0910.0245C} have found a residual age-density relationship for red sequence galaxies drawn from SDSS, which is in apparent contradiction to the results of this paper and other SDSS based stellar population studies such as \citet{Beretal06}. These two results can be reconciled, however. It is possible that the actual variation in age as a function of environment is too small for a direct detection in the present work, but visible as the residual age-density relation discussed in \citet{2009arXiv0910.0245C}. This would be consistent with the work of Clemens et al (2006, 2009)\nocite{Clemensetal06,Clemensetal09}, who find a small ($\la 0.1\;$dex) systematic offset toward older ages in high density ellipticals independently of galaxy mass.

\section{Conclusions}
We analyse the stellar population properties of 3,360 early-type galaxies as functions of their environmental density and intrinsic parameters such as velocity dispersion and dynamical mass. A major aim is to check the result found in T05 by increasing the sample size by more than a factor ten. 
The objects are morphologically selected by visual inspection from the SDSS in the narrow redshift range $0.05\leq z\leq 0.06$. The sample utilised here is part of a project called MOSES: \textbf{MO}rphologically \textbf{S}elected \textbf{E}arly-types in \textbf{S}DSS \citep[see also][]{Schawetal07b}. The most radical difference with respect to other galaxy samples constructed from SDSS is our choice of {\rm purely morphological} selection. We have {\em visually} inspected 48,023 objects in the redshift range $0.05\leq z\leq 0.1$ with $r<16.8\;$mag and divided the sample in 31,521 late-type (spiral arms, clear disc-like structures) and 16,502 early-type galaxies. The major advantage of this strategy is that our sample is not biased against star forming early-type galaxies. We calculate a 3-dimensional volume density at the location of each object, providing us with a {\em local} environmental density. Kinematics of gas and stars are determined using the code by \citet{Sarzietal06} which is based on the ppxf method by \citet{CE04}. Stellar population and emission line templates are fitted simultaneously to the galaxy spectrum, which yields clean absorption line spectra that are free from emission line contamination. Subsequently, Lick absorption line indices are measured. By means of the stellar population models of \citet{TMB03a,TMK04} the stellar population parameters age, metallicity, and \aFe\ ratio are derived using a minimised $\chi^2$ technique. Dynamical masses are calculated from a scaling between galaxy velocity dispersion and half-light radius \citep{Capetal06}.

The morphological selection algorithm is critical as it does not bias against recent star formation. We find that the distribution of ages is bimodal with a major peak at old ages and a secondary peak at young ages around $\sim 2.5\;$Gyr containing about 10 per cent of the objects. This is analogue to 'red sequence' and 'blue cloud' identified in galaxy populations usually containing both early and late type galaxies. The fraction of the young, rejuvenated early-type galaxy population depends on intrinsic galaxy properties as well as their environmental densities, i.e.\ it increases as a function of both decreasing galaxy mass and decreasing environmental density. It varies from only a few per cent for the most massive galaxies in densest regions up to 45 per cent for the smallest early-type galaxies (a few $10^{10}\;$\Msun) in the loosest environments. This implies that the impact of environment increases with decreasing galaxy mass.

The rejuvenation process detected here through optical absorption line indices may date back or stretch over several Gyrs back in time. Most likely an early-type galaxy has been rejuvenated through several minor star formation events each involving star formation of only a few per cent of the galaxy's mass. Most importantly, the rejuvenated galaxies display slightly lower \aFe\ ratios and slightly higher total metallicities. The former supports the rejuvenation scenario, as a depleted \aFe\ ratio is a clear indication for late Type~Ia supernova enrichment caused by residual star formation events. This is further supported by the fact that   the emission line spectra of most of the rejuvenated galaxies indicate the presence of minor, ongoing star formation. Some fraction of these are star forming/AGN composites. All objects that host AGN in their centres without any star formation being involved have stellar old populations. Finally, the higher metallicities suggest that  the latter does not involve purely metal-poor pristine gas but must include at least some fraction of previously enriched interstellar medium, either from internal re-processing or accretion.

The early-type galaxy population on the red sequence, instead, turns out to be insensitive to environmental densities. We find that the stellar population scaling relations including the age-mass relation are independent of the environment and only driven by galaxy mass. We confirm and statistically strengthen earlier results that luminosity weighted ages, metallicities, and \aFe\ element ratios correlate with velocity dispersion. We further show that these 'stellar population scaling relations' are well defined also for galaxy mass, which reinforces the now widely accepted conception of 'archaeological downsizing' of early-type galaxies. Most interestingly, however, these scaling relations are not sensitive to environmental densities and only driven by galaxy mass. This and the old average ages of early-type galaxies suggest that the formation of early-type galaxies is environment independent and driven only by self-regulation processes and intrinsic galaxy properties such as mass. This formation phase has lasted most of the Hubble time, but appears to have now come to an end. At lookback times of a few billion years, below $z\sim 0.2$, galaxy interactions and (minor) mergers appear to have become the major players in galaxy evolution leading to rejuvenation of mostly intermediate and low-mass early-type galaxies in low density environments.

We infer that early-type galaxy formation has undergone a phase transition a few billion years ago around $z\sim 0.2$. A self-regulated formation phase without environmental dependence has recently been superseded by a rejuvenation phase, in which the environment plays a decisive role possibly through galaxy mergers and interactions.
The universe will subsequently move into a secular evolution phase in which intrinsic galaxy properties re-gain dominance over galaxy interactions as the main driver of galaxy evolution \citep{KK04}.

The galaxy catalogue produced in this study can be found at www.icg.port.ac.uk/$\sim$thomasd.

\section*{Acknowledgements}
DT acknowledges support by grant BMBF-LPD 9901/8-111 of the Deutsche Akademie der Naturforscher Leopoldina. CM acknowldges support by the Marie Curie Excellence
Team Grant MEXT-CT-2006-042754 of the Training and
Mobility of Researchers programme financed by the European
Community. KS acknowldges support by NASA through Einstein Postdoctoral
Fellowship grant number PF9-00069 issued by the Chandra X-ray Observatory
Center, which is operated by the Smithsonian Astrophysical Observatory for and
on behalf of NASA under contract NAS8-03060. K.S. gratefully acknowledges
support from Yale University.

Funding for the SDSS and SDSS-II has been provided by the Alfred P. Sloan Foundation, the Participating Institutions, the National Science Foundation, the U.S. Department of Energy, the National Aeronautics and Space Administration, the Japanese Monbukagakusho, the Max Planck Society, and the Higher Education Funding Council for England. The SDSS Web Site is http://www.sdss.org/.

The SDSS is managed by the Astrophysical Research Consortium for the Participating Institutions. The Participating Institutions are the American Museum of Natural History, Astrophysical Institute Potsdam, University of Basel, University of Cambridge, Case Western Reserve University, University of Chicago, Drexel University, Fermilab, the Institute for Advanced Study, the Japan Participation Group, Johns Hopkins University, the Joint Institute for Nuclear Astrophysics, the Kavli Institute for Particle Astrophysics and Cosmology, the Korean Scientist Group, the Chinese Academy of Sciences (LAMOST), Los Alamos National Laboratory, the Max-Planck-Institute for Astronomy (MPIA), the Max-Planck-Institute for Astrophysics (MPA), New Mexico State University, Ohio State University, University of Pittsburgh, University of Portsmouth, Princeton University, the United States Naval Observatory, and the University of Washington.



\bsp
\label{lastpage}

\end{document}